\documentclass[sigconf]{acmart}
\AtBeginDocument{%
  }

\setcopyright{acmlicensed}

\copyrightyear{2026}
\acmYear{2026}
\setcopyright{cc}
\setcctype{by}
\acmConference[CHI '26]{Proceedings of the 2026 CHI Conference on Human Factors in Computing Systems}{April 13--17, 2026}{Barcelona, Spain}
\acmBooktitle{Proceedings of the 2026 CHI Conference on Human Factors in Computing Systems (CHI '26), April 13--17, 2026, Barcelona, Spain}
\acmDOI{10.1145/3772318.3791059}
\acmISBN{979-8-4007-2278-3/2026/04}

\usepackage{enumitem} 
\usepackage{soul} 
\usepackage{graphicx}
\usepackage{multirow}
\usepackage{booktabs}

\usepackage[utf8]{inputenc}
\usepackage{booktabs}   
\usepackage{tabularx}   
\usepackage{caption}    
\usepackage{ragged2e}   
\usepackage{setspace}   
\usepackage{fancyhdr}
\usepackage{tikz}
\usetikzlibrary{arrows.meta,positioning,calc,shapes,fit}

\usepackage{array}

\usepackage{stfloats}

\begin{document}

\title[Seeing Eye to Eye]{Seeing Eye to Eye: Enabling Cognitive Alignment Through Shared First-Person Perspective in Human-AI Collaboration}



\settopmatter{authorsperrow=4}

\author{Zhuyu Teng}
\orcid{0009-0002-9276-6101}
\affiliation{%
  \department{College of Computer Science and Technology}
  \institution{Zhejiang University}
  \city{Hangzhou}
  \country{China}
}
\email{tzhuyu@zju.edu.cn}

\author{Pei Chen}
\orcid{0000-0003-0962-6459}
\authornote{Corresponding Author}
\affiliation{%
  \department{College of Computer Science and Technology}
  \institution{Zhejiang University}
  \city{Hangzhou}
  \country{China}
}
\email{chenpei@zju.edu.cn}

\author{Yichen Cai}
\orcid{0009-0005-5422-8064}
\affiliation{%
  \department{College of Computer Science and Technology}
  \institution{Zhejiang University}
  \city{Hangzhou}
  \country{China}
}
\email{yichencai@zju.edu.cn}

\author{Ruoqing Lu}
\orcid{0009-0005-5422-8064}
\affiliation{%
  \department{College of Computer Science and Technology}
  \institution{Zhejiang University}
  \city{Hangzhou}
  \country{China}
}
\email{luruoqing@zju.edu.cn}

\author{Zhaoqu Jiang}
\orcid{0009-0006-7801-4279}
\affiliation{%
  \department{College of Computer Science and Technology}
  \institution{Zhejiang University}
  \city{Hangzhou}
  \country{China}
}
\email{zhaoqujiang@zju.edu.cn}

\author{Jiayang Li}
\orcid{0009-0008-1270-6445}
\affiliation{%
  \department{College of Computer Science and Technology}
  \institution{Zhejiang University}
  \city{Hangzhou}
  \country{China}
}
\email{jiayangli@zju.edu.cn}

\author{Weitao You}
\orcid{0000-0002-9625-5547}
\affiliation{%
  \department{College of Computer Science and Technology}
  \institution{Zhejiang University}
  \city{Hangzhou}
  \country{China}
}
\email{weitao_you@zju.edu.cn}

\author{Lingyun Sun}
\orcid{0000-0002-5561-0493}
\affiliation{%
  \department{College of Computer Science and Technology}
  \institution{Zhejiang University}
  \city{Hangzhou}
  \country{China}
}
\email{sunly@zju.edu.cn}



\renewcommand{\shortauthors}{Teng et al.}

\begin{abstract}
Despite advances in multimodal AI, current vision-based assistants often remain inefficient in collaborative tasks. We identify two key gulfs: a communication gulf, where users must translate rich parallel intentions into verbal commands due to the channel mismatch , and an understanding gulf, where AI struggles to interpret subtle embodied cues. To address these, we propose \textit{Eye2Eye}, a framework that leverages first-person perspective as a channel for human-AI cognitive alignment. It integrates three components: (1) joint attention coordination for fluid focus alignment, (2) revisable memory to maintain evolving common ground, and (3) reflective feedback allowing users to clarify and refine AI's understanding. We implement this framework in an AR prototype and evaluate it through a user study and a post-hoc pipeline evaluation. Results show that \textit{Eye2Eye} significantly reduces task completion time and interaction load while increasing trust, demonstrating its components work in concert to improve collaboration. 
\end{abstract}



\begin{teaserfigure}
  \includegraphics[width=\textwidth]{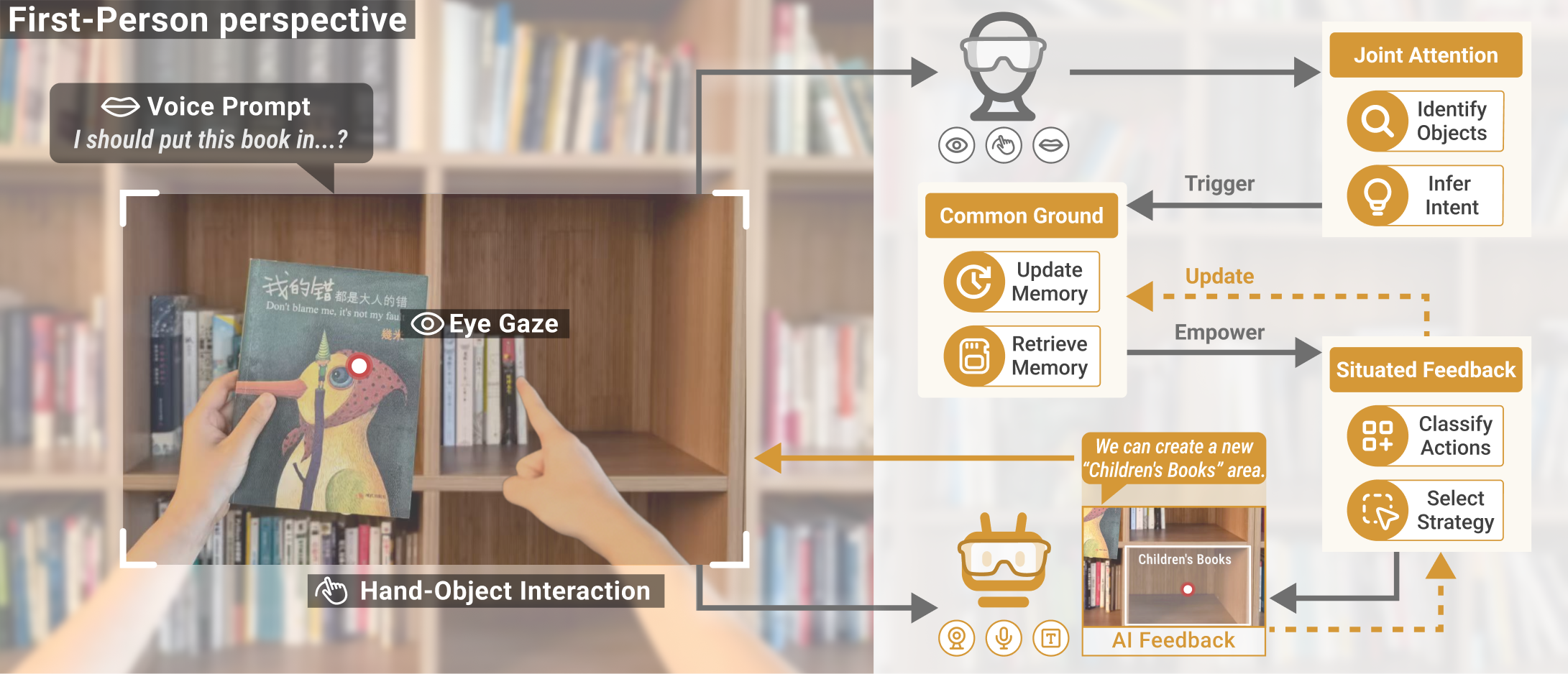}
  \caption{An example of the \textit{Eye2Eye} collaborative loop applied to a family book organization task. Wearing AR glasses, the user hesitates while deciding where to place a book in response to the AI assistant's suggestion. The AI assistant interprets implicit cues from eye gaze and hand-object interaction~(joint attention), infers intent, and consults accumulated common ground~(update and retrieve memory) to refine its understanding of the user's personalized organization rules. Based on this, it generates situated feedback through multimodal outputs~(e.g., voice prompts, visual overlays), suggesting a personalized classification~(e.g., creating a new ``Children's Books'' area), rather than suggesting generic categories. The closed loop process illustrates how \textit{Eye2Eye} transforms shared perception into shared understanding and coordinated action.}
  \Description{A composite image illustrating the "Eye2Eye" concept in a book classification task. Left: The user's first-person view through AR glasses. The user holds a book, and the system highlights the book with a bounding box and detects the user's gaze point. A voice prompt bubble shows the user asking, "I should put this book in...?". Right: A system diagram showing the collaborative loop. The loop connects the user's inputs~(Eye Gaze, Hand-Object Interaction) to three AI processing stages: (1) Joint Attention~(Identify Objects), (2) Common Ground~(Retrieve & Update Memory), and (3) Situated Feedback~(Classify Actions & Select Strategy), which then feeds back to the user via AI Feedback.}
  \label{fig:teaser}
\end{teaserfigure}


\maketitle

\section{Introduction}
An important vision for human-AI collaboration is to forge a partnership where intelligent systems and humans fluently share context, and coordinate in real time within dynamic situations~\cite{klien2005ten,seeber2020machines,shneiderman2020human}. Building on this vision,  hybrid intelligence~\cite{dellermann2019hybrid} emphasizes that humans and AI are meant to leverage their complementary strengths to improve task performance and decision accuracy~\cite{xu2023comparing}. 
Equipped with augmented reality~(AR) devices, wearable AI assistants can capture the first-person perspective and respond to users' interactions. While promising, current AI assistants remain awkward and inefficient in context-dependent tasks.


Consider a simple but common scenario: you put on AI glasses to learn how to operate a new espresso machine. When you ask, ``\textit{How do I use this?}'' the assistant provides a generic answer. But as you gaze at a panel of nearly identical buttons, you are forced into a frustrating sequence of verbal clarifications: ``\textit{What about this button?}'', ``\textit{And this one?}'' Such breakdowns are not confined to home appliances. They also appear in domains such as remote troubleshooting~\cite{gurevich2012teleadvisor} and collaborative inspection~\cite{syberfeldt2016support}, where users must repeatedly explain, re-explain, and correct the AI. These difficulties stem from two persistent gulfs that divert users from the ideal of smooth human-AI collaboration:
\begin{itemize}
    \item \textbf{The Communication Gulf:} Rooted in a fundamental channel mismatch between humans and AI~\cite{morris2025hci}. Users must ``compress'' and ``translate'' their rich, parallel physical intentions into a linear, one-dimensional linguistic stream. This constant switching between ``acting with hands'' and ``instructing by mouth'' amplifies interaction friction and incurs high grounding costs. For example, conveying a simple spatial reference may require abandoning intuitive body language in favor of cumbersome verbal descriptions like, ``the red button to the left of that switch.''
    \item \textbf{The Understanding Gulf:} The ``blindness'' of ``seeing'' in current AI visual input. While AI can recognize objects, it often fails to interpret the embodied cognition that humans naturally express in collaboration, such as a brief, hesitant gaze. Humans and AI rely on different communication channels to exchange information~\cite{rahmaniboldaji2025human,dix2003human}, and this cognitive asymmetry exacerbates the grounding problem.
\end{itemize}

These gulfs compound one another: the more effort users expend verbalizing context, the less bandwidth remains for building and maintaining alignment at the cognitive level.
Current wearable assistants often treat first-person vision as a one-way input for observation rather than a shared perception foundation for collaboration. This gap leads to a central research question: how might we exploit shared first-person perspective to achieve continuous cognitive alignment in human-AI collaboration?

In this work, we propose the \textit{Eye2Eye} framework, which elevates the first-person perspective from a passive input channel to a shared perception channel for human-AI interaction. Designed to satisfy key grounding constraints, such as copresence and visibility, the framework establishes and continuously maintains a unified cognitive foundation. To realize the vision of ``seeing eye to eye,'' our framework orchestrates human and AI perception abilities into a cohesive perceive-align-reflect loop that supports ongoing cognitive alignment:

\begin{itemize}
    \item \textbf{Joint Attention Expression:} Supporting humans and AI in fluidly establishing, shifting, and maintaining shared attention through multimodal signals such as gaze, gestures, and AR highlights.
    \item \textbf{Accumulated Common Ground:} Capturing and structuring interaction episodes into dynamic contextual memory units that persist and evolve, forming the basis for alignment over time.
    \item \textbf{Reflective Situated Feedback:} Providing context-aware, in-situ guidance (visual + verbal) based on the established common ground, while enabling users to reflect on and correct the AI's understanding.
\end{itemize}

Building on this extensible framework, we instantiated a prototype system, which leverages the first-person perspective afforded by AR glasses to support AI-guided assistance for everyday tasks that rely on contextual judgment. Through multiple rounds of pilot testing, we designed and implemented a low-cost dynamic memory module that maintains event associations and contextual understanding throughout continuous interactions, beyond treating interaction logs as static prompt inputs. The system focuses on post-response memory updates and reflection mechanisms, and enables memory coordination triggers at a lower cost than knowledge graph approaches. 

In the user study involving three task types, procedural, classificatory, and inspection, we found that multimodal expressions, including natural language, gaze, and gestures, offered unique contributions to the establishment of common ground. The superior performance of \textit{Eye2Eye} in collaboration reduced interaction friction and increased confidence. Findings from pipeline evaluation suggested the synergistic benefits of our framework as a whole.

This paper makes three primary contributions. 
\begin{enumerate}
    \renewcommand{\labelenumi}{(\arabic{enumi})}
    \item We propose the \textit{Eye2Eye} framework, which conceptualizes the first-person perspective as a shared channel for achieving cognitive alignment between humans and AI.
    \item We instantiate this framework in an AR-based prototype equipped with a real-time multimodal pipeline, demonstrating how the proposed mechanisms can be implemented in practice.
    \item We conduct a controlled user study across three types of tasks. The results show that our proposed system reduces grounding costs and interaction friction, while enhancing collaboration trust.
\end{enumerate}

\section{Related Work}

\subsection{Wearable AI Assistants for Situated Guidance}
Wearable AI assistants show significant advantages in situated task guidance compared to traditional phone-based AR systems. By embedding sensing and display capabilities directly into smart glasses, they enable real-time, hands-free support that overcomes the limited input/output and heads-down posture of smartphones~\cite{zhou2024glassmail,appel2019smartphone,weerasinghe2024real}.
Recent research has sought to expand these possibilities by leveraging natural, low-load modalities such as gaze and gesture to capture user intent and provide real-time assistance. For example, Satori~\cite{li2025satori} proactively anticipates user needs and dynamically delivers multimodal guidance. GazeNoter~\cite{tsai2025gazenoter} employs AR eye tracking to generate notes with minimal distraction, while AiGet~\cite{cai2025aiget} uses gaze detection to seamlessly embed informal learning into daily activities. Similarly, Persistent Assistant~\cite{cho2025persistent} combines gaze selection with continual task support, offering decision guidance for repetitive tasks with minimal disruption. These works illustrate how gaze and gesture can be harnessed to create more fluid interaction. 

Wearable AI assistants can also overlay digital information directly into the user's real-world view, reducing cognitive load and minimizing the need to shift attention between information sources~\cite{mostajeran2020augmented,jin2024exploring,stover2024taggar,zhang2025ieds}. Building on this principle, several systems have demonstrated benefits in learning and assembly contexts. AdapTutAR~\cite{huang2021adaptutar} provides real-time, adaptive tutoring for machine operation training, reducing user error rates. 
In complex assembly operations, Lavric et al.~\cite{lavric2021exploring} utilized low-cost 2D assets including text, images, and situational arrows, to enable efficient, in-situ authoring and delivery of AR instructions for novice workers.
Westerfield et al.~\cite{westerfield2015intelligent} combined AR with intelligent tutoring, improving efficiency in motherboard assembly through real-time recognition of user actions and error feedback. These systems show that overlaying task-relevant information in-situ can enhance learning and reduce mistakes.

Despite these advances, existing wearable AI assistants are still limited by a one-way communication model and predefined knowledge bases. They lack mechanisms to dynamically adapt to users' subjective rules, evolving goals, or implicit cues such as hesitation or gaze dwell. As a result, while effective in predefined scenarios, they fall short in achieving cognitive alignment.

\subsection{Perception from First-Person Perspective}
First-person perspective~(technically referred to as egocentric vision), captured through head-mounted cameras or AR devices, provides a direct window into a user's embodied perspective. Unlike third-person views, which often suffer from occlusion and disconnected vantage points, egocentric vision preserves the user's subjective line of sight and presents objects from a consistent viewpoint~\cite{cheng2024egothink,flegar2025first}. This perspective inherently embeds implicit information, such as gaze direction~\cite{bettadapura2015egocentric}, attention focus~\cite{grauman2024ego,jia2022egotaskqa}, and hand-object interactions~\cite{cai2016understanding,han2016ar}, which are crucial for AI to infer user intentions and task goals. Despite challenges such as limited field of view, occlusion, or motion blur~\cite{arakawa2024prism}, these properties make egocentric vision uniquely advantageous for situated collaboration.

Recent systems have started to exploit these advantages by integrating egocentric video streams with multimodal large language models~(MLLMs) to support real-time perception and reasoning. For example, Vinci~\cite{huang2024vinci} couples egocentric video with LLMs to enhance performance in first-person tasks by enabling continuous understanding and assistance. Similarly, CognitiveEMS~\cite{weerasinghe2024real} combines egocentric vision and speech input to track emergency personnel's movements in real time and deliver immediate feedback. These efforts demonstrate the potential of egocentric vision as a powerful input channel for situated AI assistance.

However, these systems primarily treat the first-person stream as a recognition input, and offer guidance based on one-shot or static perception. They do not maintain an evolving, shared context with the user. Our work goes beyond perception: we conceptualize egocentric vision as a bidirectional, shared perception space, enabling continuous cognitive alignment rather than one-time interpretation~\cite{huang2024egoexolearn}.

\begin{table*}[!t]
\small
\caption{Typical challenges in collaborating with wearable AI assistants, taking the furniture assembly task as examples.}
\label{tab:challenges}
\renewcommand{\arraystretch}{1.2}
\begin{tabularx}{\textwidth}{@{}>{\RaggedRight\arraybackslash}p{0.16\textwidth} 
                                    >{\RaggedRight\arraybackslash}p{0.22\textwidth} 
                                    X@{}}
\toprule
\textbf{Challenge Type} & \textbf{Subcategory} & \textbf{Example in Furniture Assembly} \\
\midrule

\multirow{4}{*}{Communication gulf} 
  & Attention mismatch 
  & The user points to a side panel in the corner and says ``put this here,'' but the AI assistant defaults to the central wooden board, giving incorrect guidance. \\[0.25em]

  & Over-verbalization burden 
  & The user asks ``Which screw should I use?'' and the AI assistant responds with a long explanation of screw material and threading standards, instead of simply saying ``use the 5mm screw.'' \\

\midrule

\multirow{4}{*}{Understanding gulf} 
  & Intervention blindness 
  & The user hesitates between using a drill or a screwdriver, pausing with their hand above the tools. The AI assistant fails to interpret the hesitation as a cue for timely guidance. \\[0.25em]

  & Common ground drift 
  & The user decides to mount the drawer on the left for personal convenience, but the AI assistant continues to insist on the default right-side placement specified in the manual. \\

\bottomrule
\end{tabularx}
\end{table*}

\subsection{Cognitive Alignment in Human-AI Collaboration}
Effective human-AI collaboration requires the continuous coordination of actions through communication, grounding, and feedback loops. 
Grounding in communication theory~\cite{clark1991grounding} posits that effective grounding is significantly constrained by factors like copresence and visibility. To navigate these constraints,  it is important to negotiate goals and manage attention~\cite{klein2005common, convertino2008articulating}.
Recent work highlights the role of implicit signals, such as gaze~\cite{lee2024gazepointar, chan2025improving}, body movements~\cite{zheng2022gimo} as critical cues for predicting user intent and situated disambiguation. This ongoing grounding process serves as a foundation for team cognition.
Another important theoretical perspective is that of shared mental models, which emphasizes that when team members have aligned understandings of the task and of each other’s roles, the team can perform more effectively~\cite{cannon1993shared, andrews2023role}.
Research on human-AI teaming suggests that action oriented communication helps the development of shared mental models~(SMM)~\cite{schelble2022let}. Once established, SMMs in turn help coordinate member actions and significantly enhance subsequent communication efficiency~\cite{mathieu2000influence, bierhals2007shared}.
Ultimately, this persistent shared understanding underpins distributed cognition within the human-AI team~\cite{hollan2000distributed}. By enabling the AI to actively share cognitive loads, such as error monitoring and decision support, the dyad achieves a dynamic balance, thereby improving overall collaborative efficiency.

Building on these theoretical perspectives, the practical realization of such coordination relies on cognitive alignment. It involves a continuous process in which humans and AI establish, maintain, and update a shared understanding of goals, intentions and situational context.
However, due to the differences in underlying mechanisms of information perception and expression between humans and AI~\cite{he2023interaction,mingyue2017human}, achieving cognitive alignment is challenging.
Prior research has explored frameworks for joint action and task alignment, aiming to reduce grounding costs in human-AI collaboration. 
Yin et al. introduced TaskMind, which reconstructs cognitive dependencies to model user intent behind operations~\cite{yin2025operation}. He et al. proposed a Capability-Aware Shared Mental Model that adapts through iterative clarification with humans~\cite{he2023interaction}. Yang et al. developed AMMA, which aligns human and AI task steps via state tracking and guidance planning~\cite{yang2024amma}. 
Building on this foundation, we seek to extend alignment to the shared perception, addressing the need for grounding in situated collaboration.

In this work, we propose cognitive alignment as a guiding principle for wearable AI assistants, transforming egocentric vision into a shared perception foundation for dynamic, adaptive collaboration.
Specifically, \textit{Eye2Eye} instantiates these theoretical pillars by: (1)  coordinating attention via explicit and implicit cues for grounding in communication and situated disambiguation; (2) using updatable object-card memory units to maintain a persistent shared mental model; (3) leveraging situated AR feedback to support distributed cognition within the human-AI team, enabling an efficient joint cognitive system.

\section{Concept of \textit{Eye2Eye}: A First-Person Perspective Framework for Cognitive Alignment}
Our research establishes the first-person perspective as the core shared channel for human-AI collaboration. We propose the \textit{Eye2Eye} framework as a structured approach to overcome the gulfs of communication and understanding in wearable AI interaction. This section articulates the conceptual foundations of the framework: how the motivating problems are abstracted into design requirements, and how these requirements are organized into a theory of cognitive alignment.

\subsection{Identifying Core Challenges}
To clarify the abstract problems of communication and understanding gulfs, we conducted a preliminary exploration drawing on two sources of evidence: (1) small-scale experiential trials with a wearable AI assistant, supplemented by brief informal interviews, (2) a focused synthesis of findings from related literature on situated human-AI interaction. 
In these explorations, the authors recalled concrete collaborative episodes, analyzed pain points, and reflected on situations where the AI's behavior appeared misaligned with human intentions.

From this process, we distilled four recurring challenges, organized under the two gulfs (Table~\ref{tab:challenges}). To illustrate these challenges more concretely, we take the example of a user wearing a wearable AI assistant to assemble flat-pack furniture. Within the communication gulf, we identified attention mismatch, where deixis such as ``this'' is misinterpreted and defaults to incorrect objects, and over-verbalization burden, where the AI assistant provides unnecessarily lengthy explanations that force users to overspecify their intent. Within the understanding gulf, we observed intervention blindness, where the AI assistant fails to detect confusion from implicit cues such as gaze dwell or hesitation, and common ground drift, where the AI assistant rigidly applies generic categories and overlooks user-specific reinterpretations.

While presented distinctively, these challenges are intrinsically interconnected. For example, an attention mismatch often arises from a lack of shared common ground, and over-verbalization is a compensatory behavior for the AI assistant's blindness to implicit cues. 
These challenges illustrate that current AI  assistants lack effective mechanisms for interpreting both explicit and implicit human cues, maintaining evolving shared context, and adapting feedback strategies.
Any breakdown in the cognitive alignment loop can lead to these phenomena. Therefore, we further articulate three interrelated design requirements that define how collaboration through first-person perspective should function.

\begin{figure*}[!t]
  \centering
  \includegraphics[width=0.75\textwidth]{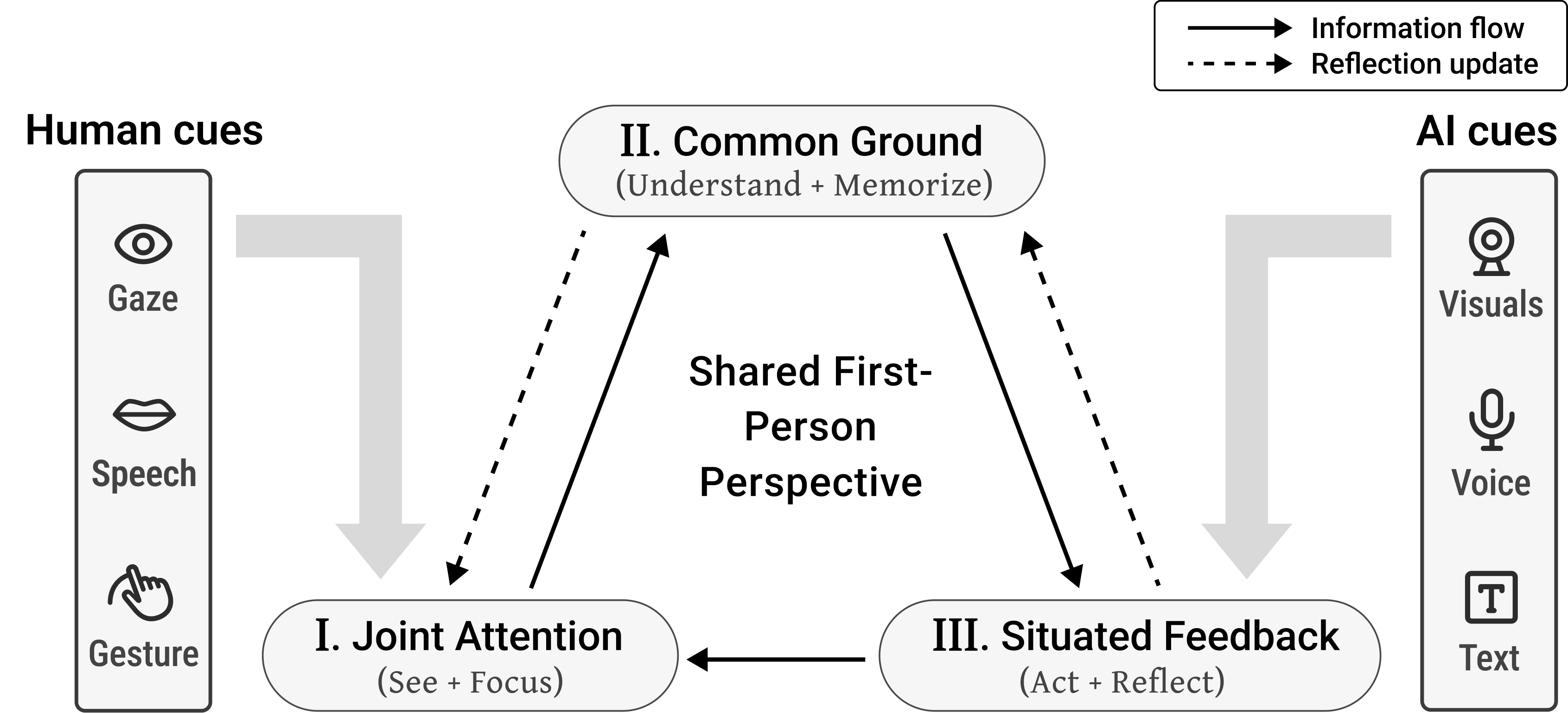}
  \caption{Framework overview of \textit{Eye2Eye}. The framework illustrates the bidirectional cognitive loop between humans~(left) and AI~(right), mediated by three core components. The process flows from (1) joint attention, aligning and interpreting the focus of attention, to (2) common ground, enriching and updating shared memory, and then to (3) situated feedback, acting and reflecting based on context. This feedback loop, in turn, updates the shared memory and can proactively generate new attention cues.}
  \Description{A triangular flow diagram representing the \textit{Eye2Eye} framework's cognitive loop centered on a Shared First-Person Perspective. It illustrates a bidirectional cycle between three components: Joint Attention, Common Ground, and Situated Feedback. Clockwise Flow: Human cues~(Gaze, Speech, Gesture) initiate the information flow~($I \rightarrow II \rightarrow III$), driving the system to perceive and act1111. Counter-Clockwise Flow: AI reflection and memory updates~($III \rightarrow II \rightarrow I$) cycle back to refine the system's understanding based on user feedback.}
  \label{fig:Framework}
\end{figure*}

\subsection{Design Requirements}

\textbf{DR1: Establish and maintain aligned attention.}
The AI assistant should ensure that humans and AI can identify, converge on, and continuously verify a shared focus of attention.  
This requires detecting and interpreting both explicit cues~(e.g., speech, pointing) and implicit cues~(e.g., gaze dwell, touch gesture), while making the AI assistant's own focus transparent through multimodal feedback.  
First-person perspective enables \textit{mutual presence and visibility}, allowing attention alignment to happen dynamically in the flow of ongoing activity.

\textbf{DR2: Establish a persistent accumulated and revisable shared understanding over time.}
The AI assistant should build a persistent and evolving common ground that integrates past interactions with the current context. 
This involves continuously accumulating shared experiences and supporting corrections, so that understanding can be revised when misconceptions are detected.  
From a first-person perspective, this means enabling \textit{persistent revisability}, allowing momentary attentional cues to evolve into durable, context-aware knowledge.

\textbf{DR3: Deliver accountable feedback.}
The AI assistant should generate feedback that is situationally appropriate, minimally disruptive, and responsive to user reactions.  
It must also monitor the outcomes of its interventions, integrating successes and failures back into the shared understanding to improve future collaboration.  
In a shared first-person view, this is supported by \textit{real-time parallelism}, enabling the system to deliver immediate and situated feedback thereby offloading cognitive burdens.

\subsection{Core Components}



As illustrated in Figure~\ref{fig:Framework}, the \textit{Eye2Eye} framework addresses the three design requirements through three interdependent components: joint attention coordination, accumulated common ground, and reflective situated feedback. The three components of the framework form a recurrent loop. Component I establishes joint attentional focus in real time; Component II sustains and revises this focus in the form of accumulated common ground; and Component III translates understanding into situated feedback while reflecting outcomes back into the common ground. This creates a dynamic cycle: I → II → III → II → I, ensuring that fleeting attentional cues evolve into persistent, revisable knowledge, and that feedback continuously refines future alignment.

\subsubsection{Component I: Joint Attention Coordination~(See + Focus)}

Leveraging the inherent real-time parallelism of shared first-person perspective, this component enables humans and AI to converge on a shared focus, and continuously align their attention within the same space.
Wearable AI assistant makes this mutual focus visible: AI can perceive the user's gaze and gestures through the headset's camera sensors, enabling it to detect the user's focus; at the same time, AI can render bounding boxes or visual highlights directly within the user's field of view, allowing the user to see precisely what AI is attending to. Such bidirectional visibility helps prevent misunderstandings and reduces the effort required to resolve ambiguous references. The design of this component is inspired by findings on shared attention in human-robot interaction~\cite{huang2011effects}. To operationalize this process, we distinguish two complementary mechanisms: how AI responds to the user's attentional cues~(from human to AI), and how it proactively communicates its own focus back to the user~(from AI to human).

\begin{figure*}[!t]
  \centering
  \includegraphics[width=\textwidth]{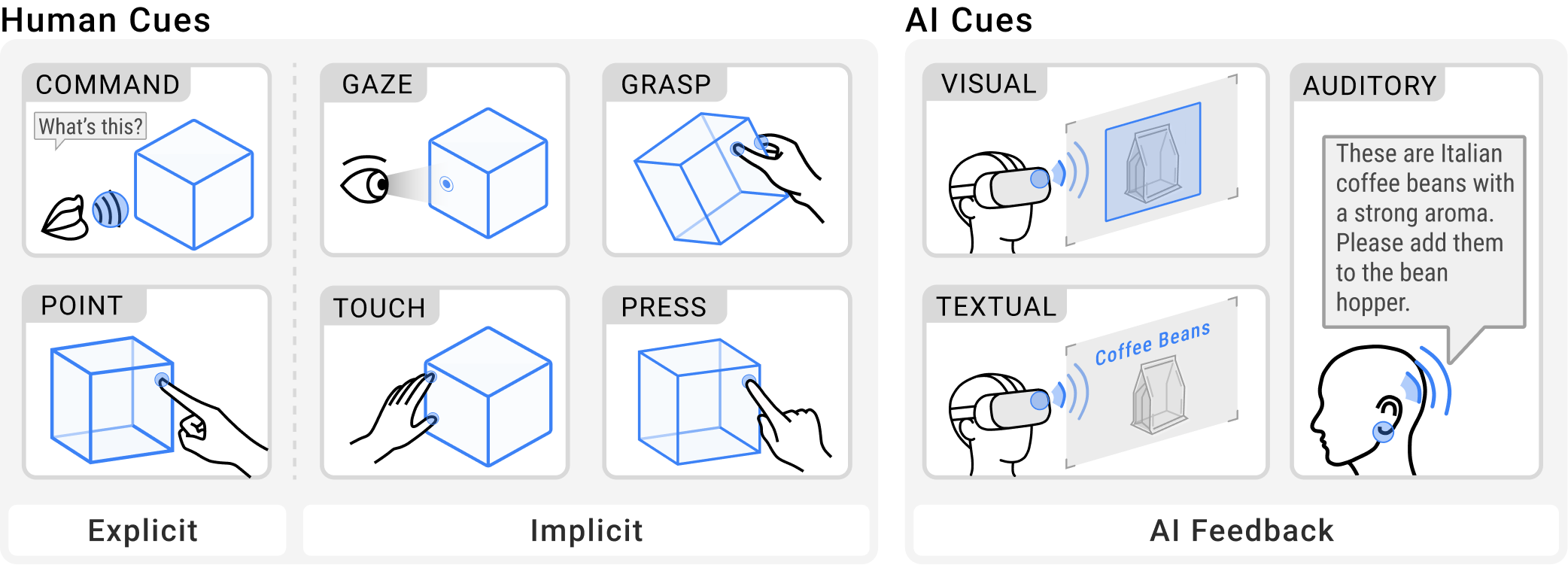}
  \caption{Bidirectional attention in a shared first-person perspective: AI captures and infers human's attention from explicit and implicit cues, subsequently aligning their attention and providing guidance through multimodal feedback.}
  \Description{A matrix of illustrations showing how attention is exchanged. Left~(Human Cues): Shows explicit cues~(User pointing, asking ``What's this?'') and implicit cues~(User gazing at an object, grasping or touching a box). Right~(AI Cues): Shows how AI responds. Visual feedback places a virtual label next to the object; Auditory feedback provides a spoken explanation (``These are Italian coffee beans...''); Textual feedback displays a floating caption~(``Coffee Beans'').}
  \label{fig:joint_attention}
\end{figure*}

\textbf{From human to AI: inferring attentional focus.}
This requires AI to interpret the user's attentional cues in order to infer their current focus. As illustrated in Figure~\ref{fig:joint_attention}, AI must account for both implicit and explicit expressions of attention.
\begin{itemize}
    \item \textbf{Implicit cues} include tracking the user's gaze to infer which object they are observing, or detecting moments of hesitation; as well as recognizing hand gestures such as touching, grasping, or pressing, which indicate the target object and the logic of the interaction. 
    \item \textbf{Explicit cues} involve the user directly guiding AI's focus through verbal commands (e.g., ``What is this?''). In addition, following the COFI definition~\cite{rezwana2023designing}, we classify deictic gestures such as pointing to an object as explicit cues, since they clearly specify the user's hand-object interaction target.
\end{itemize}

\textbf{From AI to human: communicating attentional focus.}
Once AI has inferred the user's current focus from implicit and explicit cues, maintaining alignment also requires that it reciprocally disclose its own attentional state. This reciprocal process is realized through multimodal feedback, which makes the AI's focus transparent and verifiable to the user.
\begin{itemize}
    \item \textbf{Visual expression}. 
    The AI assistant externalizes its attentional focus through spatially anchored graphical overlays, such as bounding boxes or highlights, rendered directly in the user's AR view. This provides explicit confirmation of the referenced object but requires the user's primary attentional resources.
    \item \textbf{Auditory expression}. Through spoken prompts delivered in real time, the AI communicates guidance and attentional focus without occupying the visual channel. The level of detail in these prompts can be adapted to task complexity and user expertise, supporting sequential and context-sensitive instruction.
    \item \textbf{Textual expression}. 
    Short text overlays appear in peripheral regions of the AR display, presenting object labels, system status, or concise step instructions. This modality is particularly effective in noisy environments or when the verbal channel is overloaded, as it enables quick on-demand reference whenever they wish, without interrupting the ongoing task.
\end{itemize}

\subsubsection{Component II: Accumulated Common Ground~(Understand + Memorize)}
Serving as the substrate of the framework, Component II bridges the transient alignment of Component I and the action-reflection loop of Component III, ensuring continuity across episodes.
In collaborative processes, a user's goals, attention, and rules often shift dynamically with the context. Traditional human-AI interaction tends to rely on ``one-off'' interpretations and feedback, or on treating interaction history as a static context window. However, such approaches struggle to capture the sparse evolving nature of partnership in a vast interaction history, are prone to errors in AI responses, and often force users to repeat clarifications. 
To address this fundamental issue, we introduce the principle of Accumulated Common Ground~(detailed implementation of the object card as the core unit will be introduced in Section~\ref{sec:system}). This mechanism follows two core ideas:

\textbf{Persistent accumulation.}
The system continuously accumulates shared experience by appending each interaction episode, which includes user behaviors, AI responses, and their ultimate outcomes. With each new interaction, the AI consults this shared workspace to retrieve relevant history, integrating the current situation with past context to form a rich, new understanding. This process allows the AI to act like an experienced partner, comprehending that ``this book'' is not just ``a book,'' but ``the book I hesitated over previously for my child's classification.''

\textbf{Revisability.}
Our framework acknowledges that AI's initial understanding might be incorrect. The understanding stored in each card is not a final verdict, but a continually evolving working hypothesis that is refined based on user feedback. A simple correction from the user~(e.g., ``Not this one, the one next to it'') is captured and used to refine its understanding within the common ground. This self-correcting process ensures that AI's understanding remains synchronized with the user's subjective world and becomes increasingly intelligent and reliable over time.

\subsubsection{Component III: Reflective Situated Feedback~(Act + Reflect)}

This component completes the collaborative loop by enabling AI to act on its understanding and to learn from the outcome. In a shared first-person perspective, user behavior is constantly evolving. Our goal of maintaining mutual visibility between human and AI means that it is not enough for AI to merely understand context.
Instead, AI must be able to act on its understanding and, crucially, learn from the outcomes of its actions to adapt to the user's evolving goals.
Our framework defines situated feedback as a dual mechanism, responsible for both Action and Reflection.

Component III plays a dual role.
First, user-facing action policy: Given the cognitive state of the common ground and situated interpretation of the current scene, Component III formulates an appropriate response plan to guide the user. It selects the feedback modality within the shared space so as to draw sufficient attention without undue interruption.
Second, AI-facing reflective policy: In shared perspective collaboration, AI should remain accountable for its feedback. At each turn, it monitors the user's reaction. If the user proceeds smoothly, the previous AI response is logged as a success and consolidated into memory units; otherwise, it is marked as a failure with the inferred reason for the mismatch, and the cognitive units in the common ground are refined accordingly. The outcomes of situated feedback are fed back into Component II, setting the stage for the next cycle of joint attention.


\begin{figure*}[!t]
  \centering
  \includegraphics[width=\textwidth]{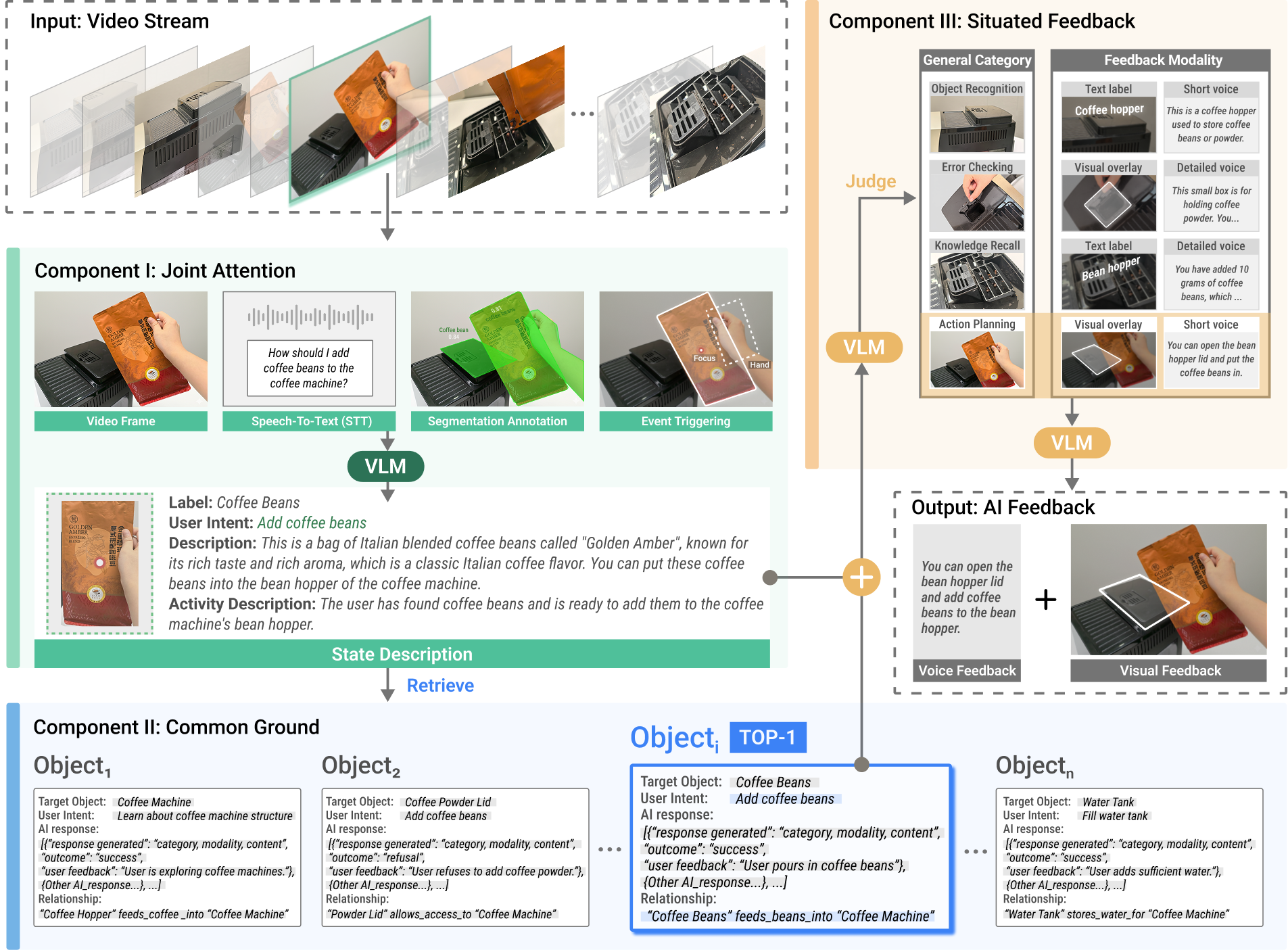}
  \caption{The technical pipeline of the implemented prototype: (1) attention trigger and interaction event understanding; (2) retrieving and revising common ground for cognitive alignment; and (3) executing the feedback strategy.}
  \Description{A technical flowchart showing the data processing pipeline from input to output. Input: A video stream showing a user holding a bag of coffee beans. Component I (Joint Attention): The system processes video frames, speech ("How should I add coffee beans?"), and segmentation masks to detect the "Coffee Beans" label and user intent. Component II (Common Ground): The system retrieves relevant "Object Cards" (e.g., Coffee Machine, Coffee Hopper) to contextualize the current state. Component III (Situated Feedback): A VLM generates a strategy (Action Planning) and outputs multimodal feedback~(Voice: "You can open the bean hopper...", Visual: highlighting the hopper lid).}
  \label{fig:Pipeline}
\end{figure*}

\section{AR Prototype System}
\label{sec:system}

Guided by the framework, we implemented an ``\textit{Eye2Eye}'' AR prototype system on Apple Vision Pro, which provides capabilities such as eye tracking, gesture sensing, and AR rendering that are essential for capturing user attention and making the AI assistant's feedback visible in real time. The prototype instantiates the theoretical components of the framework through concrete technical choices, allowing us to test how cognitive alignment can be achieved in practice. The processing pipeline of the implemented prototype is illustrated in Figure~\ref{fig:Pipeline}.

\subsection{Always-On Perception and Event-Driven Triggering (for Component I)}

\begin{figure*}[htbp]
  \centering
  \includegraphics[width=0.9\textwidth]{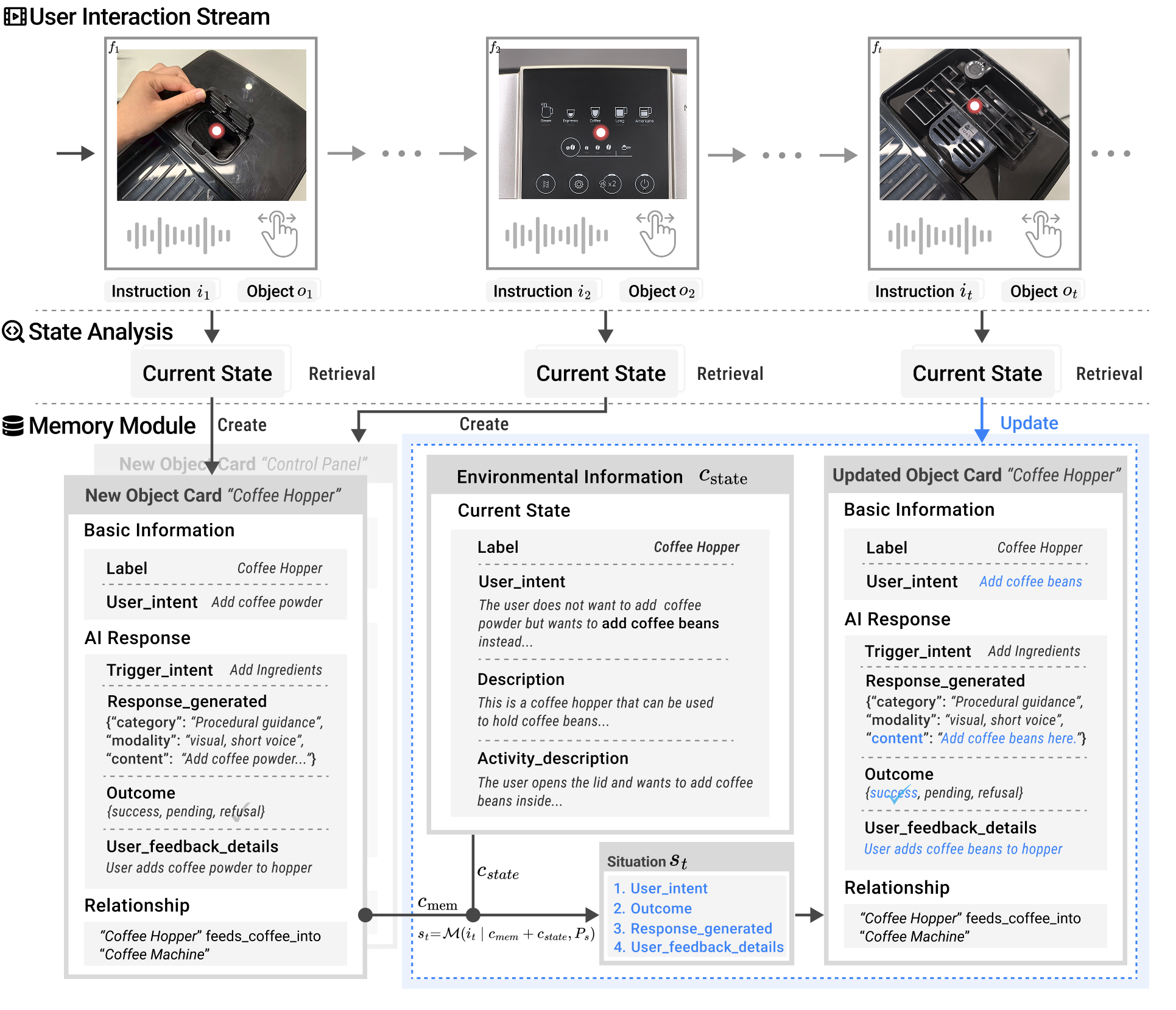}
  \caption{Details of the dynamic, self-correcting memory unit structure: each analysis of the current state~($c_{state}$) retrieves existing object cards to either create a new common ground unit or update an existing one, enabling persistent accumulation. Updates to the AI response are based on reflection of user feedback from interactions, achieving cognitive refinement.}
  \Description{A diagram detailing the "Object Card" dynamic memory structure. It shows a sequence where a new interaction updates an existing memory card. Initial State: A card for ``Coffee Hopper'' exists. Update Process: The user interacts with the hopper. The system analyzes the ``Current State'' and ``User Intent''~(changing from adding powder to adding beans). Result: The Object Card is updated with new ``User_feedback_details'' and a revised ``AI Response'' strategy, demonstrating how the system learns from correction.}
  \label{fig:common_ground}
\end{figure*}

To instantiate joint attention coordination, the prototype employs an always-on, event-driven pipeline designed to balance responsiveness and accuracy. The pipeline operates in two stages: a lightweight perception module for continuous signal capture, and a semantic interpretation module for high-level reasoning.

In the perception stage, multimodal signals are continuously collected from the Apple Vision Pro. The headset's native APIs provide real-time gaze points, while its live broadcast function streams audio and image frames that capture both objects and hand-object interactions~(e.g., touching, picking up, placing). Audio input is transcribed online through a speech-to-text~(STT) service into a textual instruction $i$. An object detection module processes the image frames $f$ to recognize key interaction events, such as prolonged gaze fixation or grasping gestures, and outputs class labels and bounding box coordinates for each recognizable object $o$ in the user's field of view. 

In the interpretation stage, high-level reasoning is triggered only when an interaction event is detected. The system engages a vision language model~(VLM). The VLM receives multimodal inputs, including the current video frame, the STT transcript, segmentation annotations, and the event type, and produces a contextualized analysis of the situation.

Through iterative testing, we determined the following attention trigger conditions to achieve effective performance: (1) when hand keypoints overlap with an object's detection box by at least 85\%, the VLM is triggered to recognize the hand-object interaction and behavior; (2) when the gaze fixation point remains on the same object for $\ge $ 6 seconds, the VLM is triggered to analyze the object of attention and interaction behavior.

\subsection{Object-Card Memory Construction and Updating~(for Component II)}


To operationalize the principle of accumulated common ground, we employed context engineering to design the memory as a set of dynamic, updatable  ``object card'' units \( \mathcal O = \{o_1, o_2, \dots\} \). Each object card serves as a contextual anchor that records how the user and the AI assistant have interacted with a specific entity. Rather than treating interaction history as a static log, this design creates a structured, updatable representation that supports both accumulation and revision of shared experiences. As shown in Figure~\ref{fig:common_ground}, a typical object card unit contains: 
label and description of the target object, inferred user intent, and their vector embeddings, AI responses with user reactions, recorded as structured entries labeled 
\textit{success}, \textit{failure}, or \textit{pending}; relationship to other objects, which form lightweight associations~(e.g., ``handle'' \textit{is\_inserted\_into} ``brew head'').

In the retrieval and updating of memory units, we adopt the idea of Retrieval-Augmented Generation (RAG).
Conventional RAG treats historical records as static prompts repeatedly fed into the VLM, which can easily overlook updates or clarifications to user intents during multi-turn interactions. To address this, and following the processing framework described earlier, we implement a trigger-retrieve-update workflow tailored to the retrieval and updating of object card units within context engineering.

When a new interaction at time $t$ occurs, the system first recognizes the current object $o_t$ and summarizes the situational state $c_{state}$ from current inputs. Then it encodes the basic information of the object into embeddings by a multimodal encoder and attempts to retrieve the most relevant object card by comparing the current object with the stored set $\mathcal O$ using cosine similarity. To ensure relevance, only candidates with similarity above a predefined threshold $\lambda$ are considered. The most similar object is selected and passed through the VLM $\mathcal M$ to obtain a retrieved context $c_{mem}$:
\begin{equation}
c_{mem} = \mathcal{M} \left( \underset{o_i \in \mathcal{O},\; \text{similarity}(o_t, o_i) > \lambda}{\arg\max} \; \text{similarity}(o_t, o_i) \right)
\end{equation}

The context $c_{mem}$ is then fused with the situational state $c_{state}$ by a fusing prompt $P_s$ to form a contextualized interpretation $s_t$: $ s_t=\mathcal M(i_t\mid c_{mem}+c_{state},P_s)$. 
After the assistant provides interpretation, the corresponding entry in the object card is updated with the interpretation $s_t$: $o_t' = \mathcal{M}(s_t\mid  o_t)$.
The outcome is then recorded, ensuring that the card evolves over time.

Common Ground functions as the cognitive nexus that bridges perception and action. As input, it receives the same multimodal information streaming from Component I, attentional foci and the detected object, which serve as “keys” for retrieving the relevant object-cards. As output, it synthesizes these foci with accumulated historical experience to generate a contextualized representation of the situation, which is then passed to Component III to guide decision-making.

\begin{figure*}[!t]
  \centering
  \includegraphics[width=0.9\textwidth]{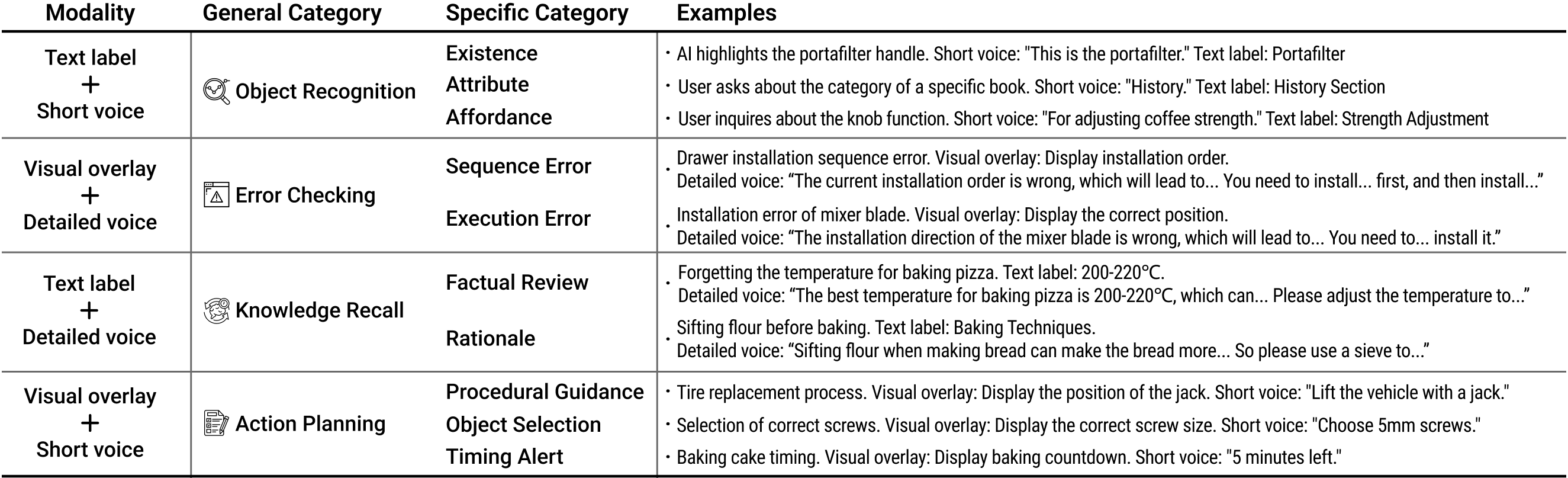}
  \caption{Mapping of situation types to feedback strategy modalities and categories, with representative examples for each case.}
  \Description{A table summarizing the mapping logic between four general situation categories and their corresponding feedback modalities. Object Recognition: Uses Text label + Short voice for quick identification~(e.g., naming attributes). Error Checking: Uses Visual overlay + Detailed voice for spatial correction and explanation~(e.g., fixing installation errors). Knowledge Recall: Uses Text label + Detailed voice for delivering factual information~(e.g., providing rationale). Action Planning: Uses Visual overlay + Short voice for immediate spatial guidance~(e.g., procedural steps). The table further details specific sub-categories and provides concrete examples for each case.}
  \label{fig:feedback}
\end{figure*}

\subsection{Multimodal Feedback Generation and Adaptation~(for Component III)}

To implement reflective situated feedback, the prototype employs a multimodal feedback generation mechanism that adapts its expression strategy to the user's attentional state and task context. The key idea is that guidance should be delivered in a way that is both transparent and minimally disruptive, ensuring that the user can immediately recognize the AI assistant's focus while continuing the task at hand.

Feedback guidance is determined based on a pre-designed feedback strategy~(Figure~\ref{fig:feedback}), which maps the current situation type to the most appropriate combination of feedback modalities.
The table organizes strategies into four general categories: object recognition, error checking, knowledge recall, and action planning.
Based on pilot study feedback, the feedback strategy follows:
(1) prioritize modalities that minimally disrupt the user's ongoing task and attention, such as \emph{action planning} feedback through \emph{visual overlay + short voice};  
(2) for essential information such as error warnings, communicate redundantly via multiple modalities (e.g., \emph{visual overlay + detailed voice});  
(3) select the modality combination best suited for the type of information. For example, for \emph{action planning} that requires clear spatial referencing, we employ \emph{visual overlay} to address ``where'' and short voice prompts to address ``what to do.'' In contrast, for \emph{knowledge recall} that requires detailed explanation, we use \emph{text labels + detailed voice}, enabling natural verbal elaboration while providing persistent textual references to reduce the user's memory load.

\subsection{Implementation Details and Performance Trade-offs}

The prototype integrates lightweight perception with high-capacity reasoning to balance responsiveness and accuracy. YOLO~\cite{Jocher_Ultralytics_YOLO_2023} is used as the instance segmentation model, enabling real-time recognition of objects and bounding boxes in the user's first-person view. We utilize Gemini 2.5 Flash for video understanding and verbal interaction. This model processes the continuous egocentric video-audio stream directly for low-latency interpretation.
For multimodal reasoning, GPT-4o~\cite{hurst2024gpt} serves as the core visual language model~(VLM) for high-level reasoning. This model processes visual, textual, and event inputs to perform complex intent inference, maintain memory and generate feedback strategies. Once GPT-4o produces textual guidance, the output is rendered into speech using Gemini, allowing the assistant's responses to be delivered in natural voice through the Apple Vision Pro's audio channel.

Performance optimization focuses on balancing latency against reasoning quality. To keep responsiveness within acceptable limits, the system employs a zero-shot inference strategy and disables chain-of-thought reasoning, maintaining an average delay of 4-5 seconds between backend response initiation and AR presentation. Pilot participants judged this latency acceptable for continuous task scenarios.
On the client side, the Apple Vision Pro streams video frames at 5 FPS, with YOLO configured at the same rate to ensure that no critical trigger events are missed. To avoid cognitive overload, the assistant processes no more than two concurrent triggers; additional triggers are discarded until the current feedback has been delivered.
On the server side, intensive inference is handled by a workstation cluster equipped with eight NVIDIA RTX 3090 GPUs, providing the computational power needed for real-time multimodal reasoning and feedback generation.

\section{User Study}
We conducted a user study to evaluate the proposed \textit{Eye2Eye} system's effects on collaboration quality,  focusing on the following research questions:
\begin{itemize}
    \item RQ1: To what extent does \textit{Eye2Eye} impact task performance and interaction in collaboration?
    \begin{itemize}
        \item RQ1.1: How does it reduce grounding costs and interaction friction?
        \item RQ1.2: How does its collaborative performance vary across different tasks?
    \end{itemize}
    \item RQ2: How does \textit{Eye2Eye} influence collaboration experience and trust between humans and AI?
    \item RQ3: In open-ended exploratory tasks, how do users view the collaborative potential and challenges with \textit{Eye2Eye}?
\end{itemize}

\subsection{Experimental Tasks}
\begin{figure*}[!t]
  \centering
  \includegraphics[width=\textwidth]{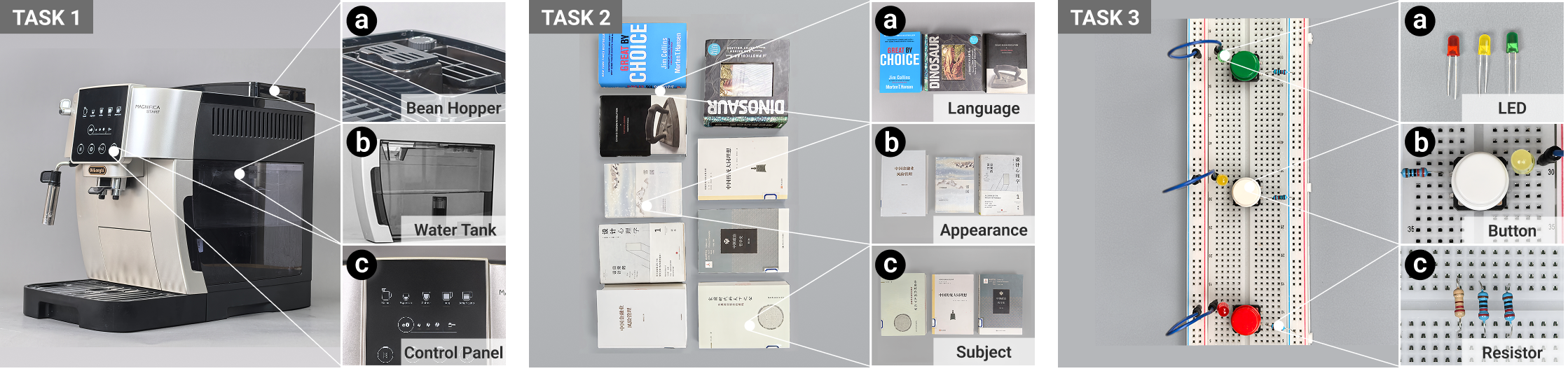}
  \caption{Experimental tasks for evaluating human-AI collaboration, including coffee machine operation, book classification, and circuit board troubleshooting.}
  \Description{A composite image displaying the three experimental tasks designed to evaluate the system. Task 1~(Left): Coffee Machine Operation. A semi-automatic espresso machine is shown with labels pointing to key components: the Bean Hopper~(top), Water Tank~(side), and Control Panel~(front). This represents a procedural task. Task 2~(Middle): Book Classification. A collection of books is laid out on a surface. Callouts indicate three classification criteria: Language~(English/Chinese), Appearance~(cover style), and Subject~(content). This represents a classification task based on subjective rules. Task 3~(Right): Circuit Board Troubleshooting. A breadboard containing electronic circuits is shown. Labels identify the components to be inspected: LED lights, Buttons, and Resistors. This represents an inspection task requiring visual search.}
  \label{fig:task}
\end{figure*}

To evaluate the robustness of \textit{Eye2Eye}, we designed three tasks with distinct cognitive and physical demands. Each task represents a category of real world activities prone to interaction friction, and targeting different aspects of human-AI collaboration (procedural memory, precise attention, and rapid classification). The difficulty levels of these tasks were controlled to be comparable (Figure~\ref{fig:task}), allowing us to verify whether \textit{Eye2Eye}'s benefits are consistent regardless of task type.
\begin{enumerate}
    \renewcommand{\labelenumi}{(\arabic{enumi})}
    \item \textit{Procedural task: coffee machine operation.} Participants were asked to use a semi-automatic espresso machine to make an Americano. This task follows a structured procedure with logical dependencies between steps, testing the system's ability to provide clear, step-by-step guidance.
    \item \textit{Classification task: book classification.} We provided 10 books of varying subjects, languages, and visual features. Participants were required to sort them based on a set of personal, subjective criteria. This task was designed to test the system's ability to build and utilize a contextual memory of user-specific rules and handle repetitive queries.
    \item \textit{Inspection task: circuit board troubleshooting.} We designed a circuit board with three parallel sub-circuits, each containing a resistor, a push button, and an LED, with one faulty component in each sub-circuit. Participants were asked to identify the faulty component in each circuit with the system's help. The task was designed to test the system's ability to guide visual search and diagnostic reasoning.
\end{enumerate}

\subsection{Procedure}
We employed a mixed experimental design: system type (\textit{Eye2Eye} vs. Baseline) was treated as a between-subjects factor, while task type~(procedural, classification, inspection) served as a within-subjects factor. This mixed design was chosen to avoid learning effects and order bias between different systems. 
The baseline was designed as a general representative of the current practice for wearable assistants, characterized by egocentric camera input coupled with large model reasoning. 
Following the comparative protocols adopted in recent studies such as ProMemAssist~\cite{pu2025promemassist} and Vinci~\cite{huang2024vinci}, we established our baseline representing the mainstream pipeline and strictly controlled for hardware and model capabilities.
Specifically, our baseline system was also implemented on the Apple Vision Pro platform, equipped with the same models but without our first-person perspective collaboration framework, ensuring a fair comparison~(Table~\ref{tab:ablation-config} in the Appendix).

The study consisted of two phases. In phase 1, participants used only one of the two systems to complete the three tasks, with task order counterbalanced to control for order effects, followed by a subjective evaluation questionnaire. In phase 2, participants were given the opportunity to freely explore the \textit{Eye2Eye} system and then took part in a semi-structured interview to provide qualitative insights into their experience. 

\subsection{Participants \& Apparatus}
We recruited 60 participants~(P01-P60, 27 female, 33 male) ages 21-28 years~($M = 22.98$, $SD = 1.66$) through social media, with 37 participants having self-reported moderate to high familiarity with AR glasses~(rating $\ge$ 3 on a 5-point scale). They were randomly assigned to two groups, each completing tasks under one of the two system conditions~(with all the three tasks). All participants had normal or corrected-to-normal vision and reported no history of vertigo or related health issues.

An Apple Vision Pro headset was used as the AR platform to provide a first-person perspective. A desktop computer hosted the backend server for data processing and system operation. Each participant carried a pocket counter and was instructed to press it whenever they perceived an error in the system's feedback, allowing us to log error occurrences. Additionally, a camera was used to record the entire interaction process for later review and analysis. All physical materials required for the tasks were provided on site.

\subsection{Measures \& Analysis}
We collected quantitative performance metrics, subjective ratings, and qualitative feedback:

\subsubsection{Objective Performance Metrics}
We compiled turn-level interaction records from timestamped system logs, task video recordings, and participants’ pocket counter entries. 
Each record includes whether the turn contained an error~(0 = correct, 1 = error), whether the user initiated a clarification action, the associated system and task condition, and timestamps. Error annotations were cross-validated between the turn-level logs and the pocket counter to ensure accuracy. 
We evaluated five task performance measures for each participant under each $system \times task$ condition: 1) task completion time, the total duration required to successfully complete a task, serving as a proxy for overall collaborative efficiency; 2) error rate, modeled as the probability of an error occurring per interaction turn~(binary outcomes: 0 = correct, 1 = error) for interaction turns; 3) clarification cost, measured by the specific subset of interaction turns where the user had to correct or request clarification following a system error; 4) interaction turns, the total number of dialogue turns required to complete the task, serving as a proxy for communication efficiency; 5) cumulative error rate to examine the temporal evolution of grounding, calculated as follows:
 $$ \text{Cumulative Error Rate}_k = \frac{\sum_{i=1}^{k} \text{Error}_i}{\sum_{i=1}^{k} \text{Turn}_i} $$

%


We analyzed all dependent variables using mixed-effects models. Linear mixed-effects models~(LME) were applied to continuous metrics that approximated a normal distribution, while generalized linear mixed-effects models~(GLMM) with appropriate link functions were used for binary and count data.
All models included System, Task, and the System$\times$Task interaction as fixed effects, with Participant included as a random intercept to account for individual variability~(model structure: $Y \sim \textit{System} \times \textit{Task} + (1 | \textit{Participant})$). 
Statistical significance of fixed effects was assessed using Type III Wald tests.
Post-hoc pairwise comparisons were performed based on estimated marginal means~(EMMs) with Bonferroni correction for multiple comparisons, and reported effect sizes and 95\% confidence intervals.



\subsubsection{Subjective User Experience Metrics}
Upon completion of each task, participants rated the perceived effectiveness of gaze, gesture, and speech as intent expression modalities, collecting 7-point Likert ratings of \textit{Eye2Eye}. 
Given the repeated measures design, we employed a linear mixed-effects Model~(LME) with task, modality, and their interaction as fixed effects, and Participant as a random intercept~(model structure: $Score \sim Task \times Modality + (1|Participant)$). Consistent with the above, we assessed main effects and conducted post-hoc pairwise comparisons.

Additionally, we used the six NASA-TLX~\cite{hart2006nasa} subscales to measure cognitive workload. 
Following prior work~\cite{chan2025improving,biocca2003networked,jian2000foundations,endsley2017toward}, we also evaluated collaboration quality with five dimensions on 7-point Likert scales: fluency, intrusiveness, copresence, shared awareness, and performance trust.
For non-repeated measures, we employed Mann-Whitney U tests due to deviations from normality, reporting the effect size ($r$).

\subsubsection{Qualitative Feedback}

After the main experimental tasks, participants engaged in a 10-minute open-ended exploration session.
They were encouraged to interact freely with the \textit{Eye2Eye} system, for example by revisiting task objects and exploring nearby rest areas~(e.g., identifying plants on the balcony, or interacting in the office pantry/lounge).
We then conducted semi-structured interviews, centered on three key dimensions: (1) participants’ experiences of a shared first-person perspective, including perceived collaborative potential and challenges; (2) how their communication strategies evolved during interaction; and (3) their overall sense of partnership with the AI.
All interviews were audio-recorded and transcribed verbatim by the first author. 
We analyzed the data using reflexive thematic analysis~\cite{braun2019thematic}, iteratively refining codes through author discussions to obtain the final themes.

\subsection{Results}

\subsubsection{RQ1: Collaboration Performance - Enhancing Efficiency and Reducing Friction}

\begin{table*}[!t]
    \small
    \centering
    \caption{Task performance metrics across conditions and task types. Values represent estimated marginal means~($EMMs$) and standard errors~($SEs$) derived from LME/GLMM~($\downarrow$ indicates lower values denote better performance).}
    \label{tab:comparison}
    \begin{tabular}{l l c c c c c c c c}
        \toprule
        \textbf{Task Type} & \textbf{Condition} & \multicolumn{2}{c}{\textbf{Completion Time(s)$\downarrow$}} & \multicolumn{2}{c}{\textbf{Error Rate(\%)$\downarrow$}} & \multicolumn{2}{c}{\textbf{Clarification Cost$\downarrow$}} & \multicolumn{2}{c}{\textbf{Interaction Turns$\downarrow$}} \\
        \cmidrule(lr){3-4} \cmidrule(lr){5-6} \cmidrule(lr){7-8} \cmidrule(lr){9-10}
        & & \textbf{$EMM$} & \textbf{$SE$} & \textbf{$EMM$} & \textbf{$SE$} & \textbf{$EMM$} & \textbf{$SE$} & \textbf{$EMM$} & \textbf{$SE$} \\
        \midrule
        \textbf{Procedural} & Baseline & 430.00 & 29.90 & 11.02 & 1.47 & 1.27 & 0.22 & 16.34 & 0.97 \\
        & System & 365.00 & 24.70 & 5.07 & 1.15 & 0.87 & 0.18 & 12.54 & 0.80 \\
        \addlinespace
        \textbf{Classification} & Baseline & 309.00 & 20.90 & 14.50 & 1.90 & 1.68 & 0.25 & 12.76 & 0.81 \\
        & System & 354.00 & 23.90 & 5.10 & 1.17 & 0.55 & 1.34 & 12.34 & 0.79 \\
        \addlinespace
        \textbf{Inspection} & Baseline & 310.00 & 20.90 & 7.39 & 1.55 & 0.64 & 0.11 & 9.83 & 0.68 \\
        & System & 278.00 & 18.80 & 3.47 & 1.11 & 0.36 & 0.21 & 9.05 & 0.65 \\
        \textbf{Overall} & Baseline & 345.00 & 13.70 & 10.62 & 1.07 & 1.11 & 0.13 & 12.70 & 0.62 \\
        & System & 330.00 & 13.10 & 4.48 & 0.71 & 0.57 & 0.09 & 11.20 & 0.56 \\
        \bottomrule
    \end{tabular}
\end{table*}

\begin{figure*}[!t]
  \centering
  \includegraphics[width=\textwidth]{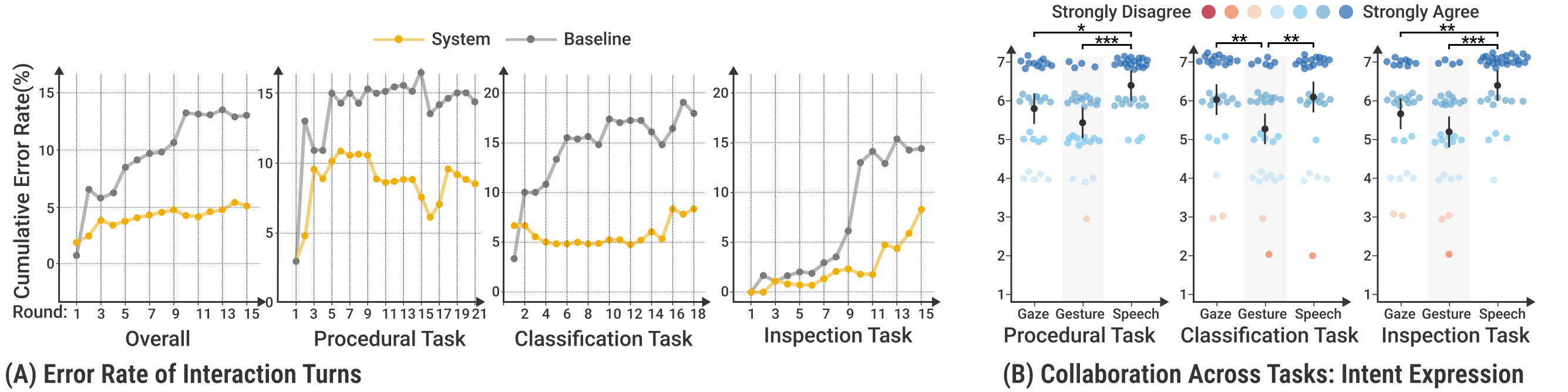}
  \caption{Collaborative performance comparison between \textit{Eye2Eye} and the baseline system assessed using linear mixed-effects models. (A) Growth curves of cumulative error rates over interaction rounds. (B) Subjective ratings of intent expression modalities~(gaze, gesture, speech), showing estimated marginal means~(EMMs) and 95\% CI.}
  \Description{(A) Line Charts~(Error Rate): Three charts showing "Cumulative Error Rate" over interaction rounds. The "System" line (Eye2Eye, yellow) remains significantly lower and flatter than the "Baseline" line (gray) across Overall, Procedural, and Inspection tasks, indicating \textit{Eye2Eye} prevents error accumulation. (B) Dot Plots~(Modality Preference): User ratings for Gaze, Gesture, and Speech. Speech is consistently rated highest~(most effective) across all tasks. Gaze and Gesture show wider variance, with Gaze being rated highly in the Classification task.}
  \label{fig:performance}
\end{figure*}


We obtained 2,338 turn-level interaction records, and Table~\ref{tab:comparison} summarizes the mixed-effects model estimates~(EMMs and SEs) for each metric across the three tasks and overall system performance.

Overall, \textit{Eye2Eye} improved accuracy and reduced grounding friction. We found a significant main effect of system on error rate~($\chi^2(1) = 22.94$, $p < .001$). Specifically, \textit{Eye2Eye} reduced the likelihood of errors by approximately 58\% compared to the baseline~($\beta = -0.93$, 95\% CI $[-1.31, -0.55]$, $p < .001$).
Similarly, for clarification cost, we observed a significant main effect of system~($\chi^2(1) = 13.80$, $p < .001$), with \textit{Eye2Eye} reducing clarification turns by approximately 50\%~($\beta = -0.69$,  95\% CI $[-1.06, -0.32]$, $p < .001$). These results indicate a robust improvement in collaboration quality regardless of task type.

For interaction efficiency and stability, we explored potential trends suggesting that the benefits of the system might be task dependent.
Significant system $\times$ task interactions were observed for both interaction turns~($\chi^2(2) = 6.08$, $p = .048$) and cumulative error rate~($F(2, 2189.20) = 16.24$, $p < .001$). Post-hoc comparisons revealed that the efficiency gain was most pronounced in the procedural task, where \textit{Eye2Eye} reduced interaction turns by 23\% compared to the baseline~($\beta = -0.26$, 95\% CI $[-0.44, -0.09]$, $p = .002$). 
As shown by the growth curves in Figure~\ref{fig:performance}.A, \textit{Eye2Eye} effectively mitigated error propagation over time, achieving reductions in cumulative error rates of 8.6\% in the classification task~($\beta = -0.086$, 95\% CI $[-0.116, -0.056]$, $p < .001$) and 4.1\% in the procedural task~($\beta = -0.041$, 95\% CI $[-0.071, -0.011]$, $p = .007$).

Regarding temporal efficiency, neither the main effect of system nor the system $\times$ task interaction reached statistical significance for task completion time.
However, the EMMs in Table~\ref{tab:comparison} suggest a distinct trend: \textit{Eye2Eye} reduced completion time on average~($EMM=330$s vs. $EMM=345$s), but required longer completion time in the classification task~($EMM=354$s vs. $EMM=309$s).
As detailed in the qualitative findings~(section 5.5.3), this likely reflects the ''micro-interruption costs'' of proactive assistance in subjective decision-making tasks.

Additionally, we assessed the perceived effectiveness of gaze, gesture, and speech to understand performance differences across tasks~(Figure~\ref{fig:performance}.B).
LME analysis revealed a highly significant main effect of modality~($F(2, 232) = 25.46$, $p < .001$), but the task$\times$modality interaction was not statistically significant.
In the inspection and procedural tasks, speech was rated significantly higher than both gaze and gesture.
In the classification task, however, we found no significant difference between speech and gaze, both significantly higher than gesture~(speech - gesture: $\beta = 0.83$, 95\% CI $[0.25, 1.42]$, $p = .002$; gaze - gesture: $\beta = 0.77$, 95\% CI $[0.18, 1.35]$, $p = .005$).
Qualitative findings provide further insights into these results.

\subsubsection{RQ2: Collaboration Experience - with Lower Load and Higher Trust}
\begin{figure*}[!t]
  \centering
  \includegraphics[width=\textwidth]{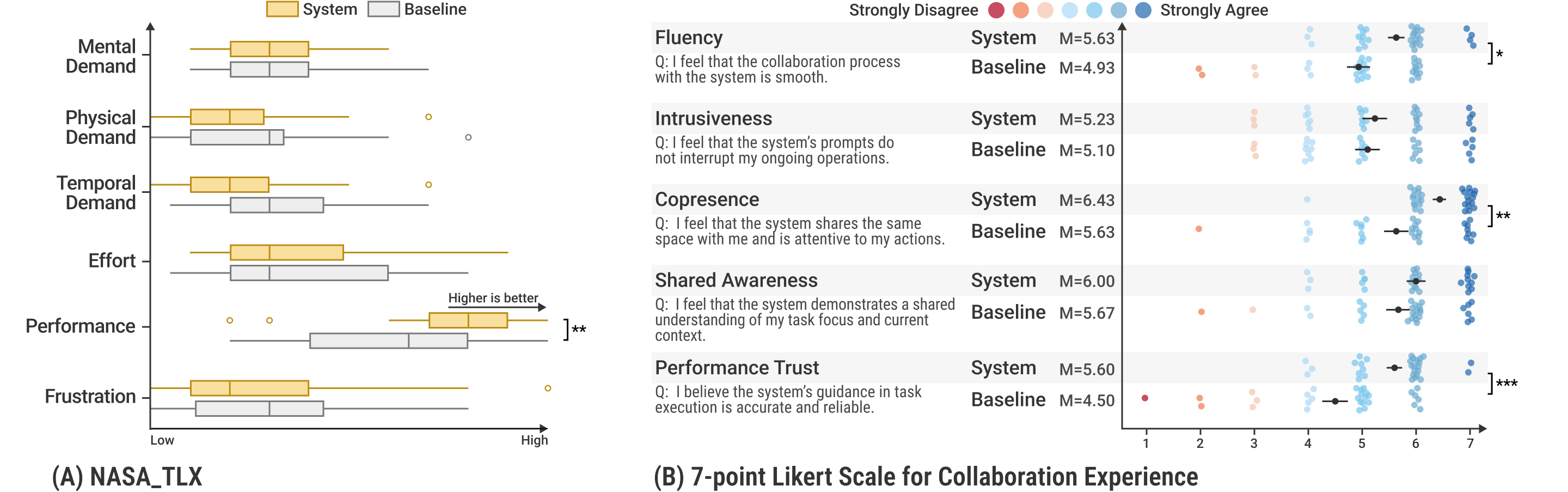}
  \caption{Subjective ratings of collaboration experience assessed using Mann-Whitney U tests. (A) Boxplots of NASA-TLX workload scores, showing medians, interquartile ranges~(IQR), and whiskers~(outliers shown as points). (B) Distributions of individual responses on five collaboration measures~(7-point Likert scale: fluency, intrusiveness, copresence, shared awareness, performance trust), with mean values ± SE indicated.}
  \Description{(A) Boxplots (NASA-TLX): Comparison of workload scores. The "System" (yellow) shows lower medians (better) than "Baseline" (gray) for Mental Demand, Physical Demand, Temporal Demand, Effort, and Frustration. Performance score is higher for the System. (B) Dot Plots (Collaboration Quality): 7-point Likert scale ratings. The "System" (blue dots) shows a strong shift towards "Strongly Agree" (scores 6-7) for Copresence, Shared Awareness, and Performance Trust compared to the Baseline.}
  \label{fig:experience}
\end{figure*}

According to the results of NASA-TLX ratings~(Figure~\ref{fig:experience}.A), \textit{Eye2Eye} had a significant effect~($r=0.39$, $p= .023$) on performance~($M = 76.83$ vs. $M = 60.83$). What's more, compared to baseline, \textit{Eye2Eye} had lower mental demand~($M = 30.83$ vs. $M = 35.50$), physical demand~($M = 21.17$ vs. $M = 36.33$), temporal demand~($M = 22.83$ vs. $M = 31.17$), effort~($M = 34.67$ vs. $M = 37.67$) and frustration~($M = 26.17$ vs. $M = 31.83$).

About collaboration experience~(Figure~\ref{fig:experience}.B), the results revealed that \textit{Eye2Eye} achieved significantly higher scores in fluency~($r = .28$, $p = .022$), copresence~($r = .35$, $p = .004$) and performance trust~($r = .45$, $p < .001$). This indicates that our framework successfully fostered the perception of the AI as a present, reliable, and understanding partner, thereby effectively bridging the critical communication gulf and understanding gulf.
In addition, participants reported that \textit{Eye2Eye} did not impose additional intrusiveness~($M =5.23$, $SD =1.25$ vs. $M = 5.10$, $SD = 1.26$) and exhibited stronger shared awareness~($M =6.00$, $SD =0.95$ vs. $M = 5.67$, $SD = 1.18$) of the current shared visual perspective.




\subsubsection{RQ3: Shifts in Collaboration - Potential and Challenges}

We identified five core themes based on interview and observations: visual copresence, partnership, cognitive pacing, intent expression, and task contingency. These themes reflect participants' perceptions and their collaborative behavior adaptations.

\textbf{Visual copresence as a bridge for grounding.}
Participants consistently reported that \textit{Eye2Eye} improved communication efficiency. A behavioral shift was observed: participants minimized verbal descriptions, adopting deictic referencing~(e.g., simply looking or pointing) to verify cognitive alignment.
P42 highlighted the sense of copresence: ``\textit{I truly felt we were in a same collaborative space... Communicating through gaze and gestures in that same space allowed us to really understand each other, so we were able to handle professional tasks.}'' This efficiency was particularly noted in repetitive tasks, ``\textit{Especially in the book task, there's a lot of stuff to look at. Eye gaze is great for focusing. The benefit is that when there are a lot of repetitive objects, selecting with your eyes is much more accurate}''~(P36).
The ``what you see is what you get'' interaction enabled both the user and AI to attend to the same specific detail, rather than a vague scene. 

However, several participants~(P11, P16, P27, P35, P41) reported that while the system understood general intent, it struggled with micro structures.
P16 pointed out limitations in fine-grained visual understanding during the inspection task: ``\textit{It can tell me this is a power module, but it can't always accurately identify the polarity.}''

\textbf{Partnership and trust.}
Participants noted proactive cues such as highlights and textual labels communicated the AI's moment-by-moment attention, reasoning, and internal state~(P9, P13, P26, P32, P47, P43, P49). Knowing the AI's reasoning basis reduced redundant checking behaviors.
``\textit{In the first system~(baseline), I didn't know if the AI understood... so I kept rotating the circuit board to show it. It never expressed what it saw. The second system~(Eye2Eye) improved this opacity significantly}''~(P47).
Crucially, the accumulated common ground was identified as a key driver of trust. Participants reported a deeper sense of consensus, noting that the system's ability to internalize user-specific rules and maintain continuity across tasks reduced misunderstandings~(P29, P34, P37, P47).
``\textit{It remembered my logic. It didn't just give me the generic biology category anymore; instead, it sorted it into the illustrated books section I set up earlier.}''~(P29).

However, this proactive partnership introduced a nuance regarding sense of agency. 
While most appreciated the support, P22 noted a preference for the baseline's passive control in specific cases: ``\textit{If I'm just going for a walk, it~(Eye2Eye) helps by offering information without making me think. However, when I have a specific target, (baseline's) voice control gives me a better sense of control.}''

\textbf{The cognitive pace in proactive collaboration.}
Although proactive guidance was generally welcomed, it raised specific concerns regarding interruptions caused by system latency~(P6, P15, P16, P27, P36).
First, intent recognition sometimes struggled to keep pace with users' shifting thoughts. 
This was most pronounced in the classification task. The system's 4-5s processing latency often lagged behind these quick decisions, creating ``micro-interruptions'' rather than assistance.
P15 noted: ``\textit{It tried to help, but sometimes I had already made up my mind before the highlight appeared. In those moments, the `help' felt like a distraction.}'' P16 similarly noted that during unexpected events, such as  when a sudden physical interruption like spilling a cup of coffee caused the user to shift focus, the system relied on outdated memory.

Furthermore, the mechanism of gaze and gesture alignment occasionally interfered with natural viewing. P16 and P36 reported that the system's aggressive attempts to capture attention targets could visually obstruct the workspace or break their focus.
Several participants reported intentionally slowing down physical movements or maintaining a static ``gaze dwell'' to allow the AI sufficient time to process and render guidance, effectively recalibrating their own pace to match the system~(P6, P27).

\textbf{Explicit and implicit cues of intent expression.}
Participants explained the rationale behind the high ratings for speech~(RQ1): because verbal communication is always the most direct way to convey intent, making its role in AR assistants irreplaceable.
Despite this, participants strongly affirmed the necessity of multimodal interaction, highlighting the unique affordances of implicit cues.
P9 reported, ``\textit{Repeating verbal commands over and over is definitely inefficient and frustrating.}''

Gestures were perceived as highly precise for spatial referencing, with P10 noting that they ``\textit{help with fine manipulation and handling multiple parts.}'' However, ambiguity remained regarding intent inference. As P11 explained: ``\textit{I'm holding something doesn't mean I want to use it. I might just be holding it for nothing.}''
Gaze was widely praised as an effortless and natural communication channel. P15 appreciated that ``\textit{looking signals interest without needing a complex background explanation.}'' Yet, this modality suffered from instability. P21 pointed out that ``\textit{unintentional gaze drift or getting distracted can trigger false recognition.}'' In summary, both implicit cues face a common challenge that the system often misinterprets meaningless or task-irrelevant behaviors as intentional input.

Participants emphasized the scenario of communication failure recovery. P43 and P55 reported that when implicit inference failed or ambiguity arose, they reverted to explicit verbal commands to rectify the system's understanding. P55 described a repair strategy where gestures establish the target, followed by explicit verbal instructions for targeted correction or detailed refinement.

\textbf{Task-dependent collaborative needs.}
Participants emphasized that task contingency plays a key role in shaping how \textit{Eye2Eye} supports collaboration. 
For procedural tasks, users valued situated feedback and expected the AI to proactively provide timely visual or verbal guidance based on the previous step, as recorded by the memory.
``\textit{If the system sees I messed up a step, it needs to tell me how to fix it right away}''~(P22).
For classification tasks, users emphasized the AI's ability to remember and apply personalized rules, highlighting the importance of the memory module.
``\textit{At the very least, don't make me ask for the rules again for every single book}''~(P5).
For inspection tasks, users prioritized precise visual attention alignment. AI's ability to assist in observation and localization benefited from a bidirectional attention coordination, both understanding the user's gaze and actively guiding it.
``\textit{It needs to understand what I just touched and tell me if I did it right}''~(P3).

\begin{figure*}[!t]
  \centering
  \includegraphics[width=\textwidth]{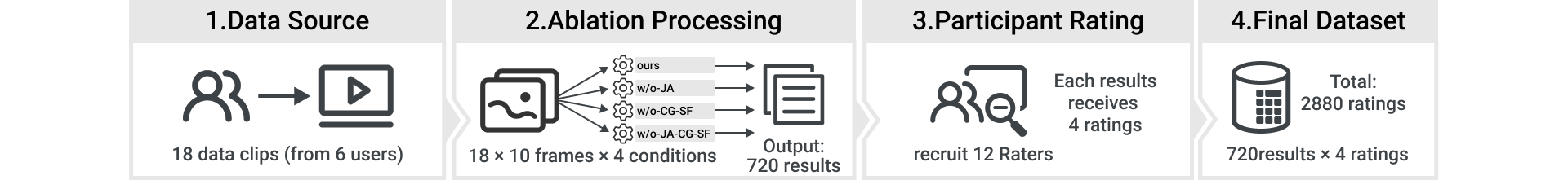}
  \caption{Data preparation and evaluation workflow for the offline ablation study.}
  \Description{The pipeline illustrates how 18 source clips from the user study were processed across four conditions to generate 720 outputs, which were then evaluated by 12 raters to yield a final dataset of 2,880 ratings.}
  \label{fig:data}
\end{figure*}

\section{Post-hoc Pipeline Evaluation}

We conducted a post-hoc evaluation of the \textit{Eye2Eye} guided pipeline to assess whether its components exhibit strong synergy and go beyond simple modular stacking.
Using real, continuous human–AI interaction data from user study, we implemented the ablation study via an offline simulation for two main purposes. First, it maintains the continuous interaction context required for the ablated variants, for example Component III still can function and update its state based on user cues. Second, although it prevents real-time user reaction, this trade-off ensures consistent input across conditions. This isolation allows us to evaluate the intrinsic performance of the components without the confounding effects of variable user adaptation.
Overall, results show \textit{Eye2Eye} effectively leverages the shared first-person perspective to outperform baselines.


\subsection{Experimental Setups}
\subsubsection{Condition Setups}

We designed three ablated variants to compare against the full \textit{Eye2Eye} framework. These ablated models were designed to leverage different aspects of the first-person perspective, but not the full extent of our framework. 
First, \textbf{pipeline w/o-JA} removes the joint attention alignment process in order to measure its individual contribution to understanding alignment.
Second, \textbf{pipeline w/o-CG-SF} removes both the cumulative grounding and corrective feedback processes~(component 2 and component 3) due to their interdependency. It is used to evaluate the benefit of maintaining and updating a shared common ground in collaboration.
Third, \textbf{pipeline w/o-JA-CG-SF} uses GPT-4o alone without any \textit{Eye2Eye} specific components, serving as a model-only baseline.
To ensure a fair comparison, all baselines were kept as consistent as possible with the full \textit{Eye2Eye} system, with similar prompt structure and same VLM. Table~\ref{tab:ablation-config} summarizes the detailed configurations.

\subsubsection{Study Design \& Data Preparation}


To evaluate using real-world data, we sampled interaction segments from six participants across three tasks, extracting continuous 10-frame clips for each. We adopted a unified output format for each condition, including object recognition, analysis of intent and activities, and guidance suggestions. 

Using a within-subjects design, we recruited 12 participants to rate the accuracy and usefulness of the output  on a 7-point Likert scale.
Each participant reviewed these egocentric clips alongside the corresponding outputs, assessing 6 randomly selected interaction data clips with the continuous 10-frame context. 
As shown in Figure~\ref{fig:data}, the study yielded 180 evaluation samples and collected 2,880 ratings~(6 sources $\times$ 3 tasks $\times$ 4 system conditions $\times$ 10 frames $\times$ 4 raters).
Data were analyzed using linear mixed-effects model~(LME) for the repeated measures, specifying condition as a fixed effect, participant and clip as random intercepts~(model structure: $Score \sim Condition + (1|Participant) + (1|Clip)$).

\subsection{Results}

\begin{figure*}[!t]
  \centering
  \includegraphics[width=\textwidth]{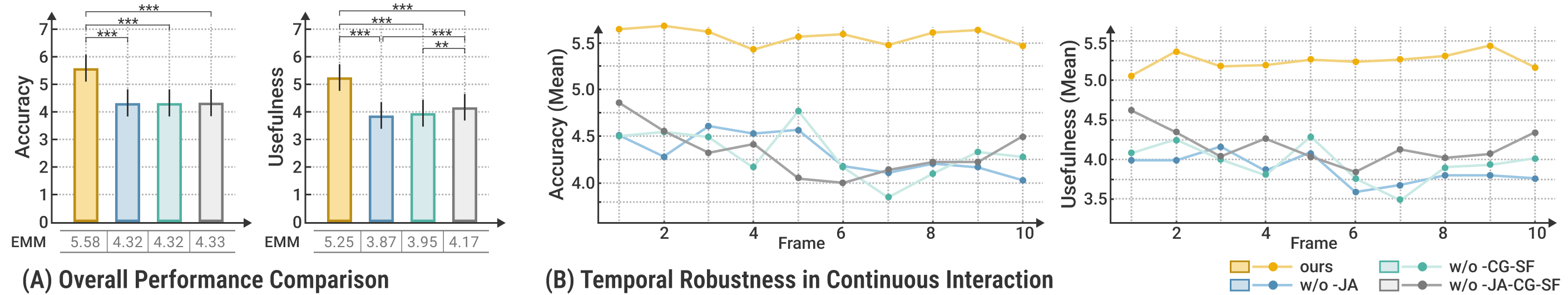}
  \caption{Ratings for the accuracy and usefulness across different pipeline conditions. (A) Overall accuracy and usefulness ratings with estimated marginal means~(EMMs) and 95\% CI. (B) Temporal robustness trends over interaction with mean scores.}
  \Description{(A) Bar Charts: Comparison of Accuracy and Usefulness across four conditions: Ours, w/o-JA (No Joint Attention), w/o-CG-SF (No Common Ground/Feedback), and Baseline. "Ours" (yellow bar) is significantly higher than all other conditions, which are roughly equal. (B) Line Charts: Temporal robustness. The "Ours" line maintains high accuracy and usefulness over time (frames 1-10), while all ablated versions show a declining trend, proving the stability of the full framework.}
  \label{fig:eva_results}
\end{figure*}

As shown in Figure~\ref{fig:eva_results}.A, for accuracy, we found a significant main effect of condition~($F(3, 2842.98) = 165.30$, $p < .001$). 
Post-hoc comparisons revealed that the full \textit{Eye2Eye} pipeline achieved the highest accuracy~($EMM = 5.58$, $SE = 0.23$), significantly outperforming the w/o-JA, w/o-CG-SF, and w/o-JA-CG-SF by margins of $\beta = 1.426$, $1.25$, and $1.25$ respectively~($p < .001$ for all).
Crucially, no significant differences were observed among the three baseline conditions. This suggests that the pipeline components are highly synergistic, and removing any single module degrades accuracy directly to the level of the raw model.
For usefulness~(Figure~\ref{fig:eva_results}.B), results showed a similar significant main effect~($F(3, 2403.02) = 168.35$, $p < .001$). 
The full \textit{Eye2Eye} pipeline achieved the highest ratings~($EMM = 5.25$, $SE = 0.22$), significantly outperforming all baseline conditions~($p < .001$ for all comparisons).
Interestingly, comparisons among baselines revealed a significant performance drop when using partial pipelines. The w/o-JA-CG-SF~($EMM = 4.17$) was rated significantly higher than the w/o-JA~($\beta = 0.30, p < .001$) and w/o-CG-SF~($\beta = 0.220, p = .01$).
This indicates that incomplete mechanisms may introduce confusion and can be detrimental to perceived usefulness.

In terms of continuous context performance~(Figure~\ref{fig:eva_results}.B), the \textit{Eye2Eye} pipeline maintained higher and more stable scores for both accuracy and usefulness. In contrast, the other pipelines showed a notable decline in performance as the interaction progressed.
These findings further support that the synergy in \textit{Eye2Eye} helps maintain consistent and effective understanding and guidance for long periods.

\section{Discussion}
\subsection{Reflection of Framework: From Visual Copresence to Joint Cognition}

\textbf{Evidence for bridging gulfs through visual copresence.}
Collaborating through a shared perspective can enhance the AI’s situational awareness~\cite{johnson2023unmapped,salikutluk2024evaluation}.
Specifically, the significant reduction in error rates and clarification costs~(RQ1) serves as empirical evidence that \textit{Eye2Eye} effectively lowers grounding costs. This confirms that our framework successfully satisfies the grounding constraints of visibility, copresence, and simultaneity~\cite{clark1991grounding}.
The efficient visual deixis of implicit embodied cues in this human-AI shared vision effectively bridges the communication gulf.

Additionally, clear feedback and revisable consensus, maintained the shared mental model during collaboration, effectively bridging the gulf of understanding. Results indicate that as the conversation progressed, the memory units of the object cards significantly enhanced system trust and transparency. Meanwhile, \textit{Eye2Eye} understands and responds to users' embodied behaviors like their gaze and gestures, frees up cognitive resources. 
This finding aligns with prior work demonstrating that utilizing multimodal embodied signals as grounding behaviors fosters more effective collaboration~\cite{mundy2007attention, zhang2025prompting,andrist2017looking}.
This responsiveness contributed to further narrowing the cognitive gap between human and AI in subsequent interactions.

\textbf{Towards a human–AI joint cognitive system.}
Although the HCI community widely advocates for treating AI as a partner rather than a tool~\cite{shi2023understanding,schmidt2023interacting}, our concept of the shared first-person perspective cognitive alignment step towards the vision of a joint cognitive system~(JCS)~\cite{dalal1994design}.
This transformation is evidenced by the dynamic distribution of cognitive tasks between the human and the AI~\cite{hoc2001towards,zhou2025examining,xu2024intuitive}. Our experiments clearly revealed this ``division of cognitive labor'': for example, in the classification task, the user was responsible for high-level, subjective rule-making, while the AI undertook the laborious tasks of memory and rule application; in the circuit board inspection, the user provided flexible visual search, while the AI was responsible for precise pattern matching and localization. In the baseline condition, by contrast, the human was the sole decision-maker and actor, with the AI serving only as a passive, on-demand information resource. Therefore, the performance gains are not merely the result of leveraging a better tool, but rather the superior efficiency achieved by the JCS.

Furthermore, the role of the AI becomes more flexible, proactive, and multifaceted. Within the \textit{Eye2Eye} framework, the AI is no longer a simple facilitator or generator, but shares responsibility for cognitive tasks such as state monitoring and next-step planning. 
The control is dynamically distributed between human and AI, blurring the boundaries of a pure active or passive relationship. Correspondingly, users' role shifts from that of a pure commander to a collaborator, because they must also confirm and correct the AI's understanding.


\subsection{Design Implications for First-Person Perspective Cognitive Alignment}

While \textit{Eye2Eye} demonstrates a positive exploration of cognitive alignment via the first-person perspective, our study also reveals several unexpected design tensions.
Careful consideration of these tensions is critical for developing flexible and helpful egocentric AI assistants.

\textbf{Tensions of proactivity: the need for strategic silence.}
While proactive assistance helps cognitive offloading, it also introduced a critical tension between helpful guidance and micro-interruptions. This trade-off explains the quantitative ``fewer rounds but longer time'' phenomenon in the classification task~(Table~\ref{tab:comparison}).
Future assistants need to adopt strategic silence, for example, modeling working memory load to determine the optimal timing for intervention~\cite{pu2025promemassist}.


\textbf{Tensions of attention: gaze as aid vs. gaze as disruption.}
It is observed that gaze could be both beneficial and disruptive from human to AI or from AI to human.
About the gaze from human to AI, when the AI assistant accurately captured and interpreted the user's gaze it helped quickly locate intended targets, reduce verbal description. When gaze was misinterpreted, it could lead to incorrect focus, repeated task interruptions, and even communication breakdowns. 
About the gaze from AI to human, well-placed visual overlays guided primary attention effectively and complemented text for deferred reference and voice for efficient, hands-free communication in visually complex tasks. Conversely, poorly positioned overlays could obscure critical environmental cues, or compete with real-world focal points.  
Participants highlighted the need for clear, low-effort mechanisms to manage gaze interaction in both directions. Such mechanisms would enable users to quickly correct gaze interpretation while also allowing them to toggle or adapt the layout of feedback modalities~\cite{li2024situationadapt}.

\textbf{Tensions of explicit/implicit cues: spatial intuition vs. semantic precision.}
Our qualitative results indicate that implicit cues facilitate effortless spatial referencing but struggle with precise intent inference. Conversely, explicit speech excels at intent definition, yet incurs high cognitive effort and referential ambiguity when describing complex spatial targets.
This resonates with the HCI community's long-standing interest in multimodal interaction for improving naturalness, robustness and expressiveness~\cite{oviatt2000perceptual}.
Effective future interfaces must orchestrate these explicit and implicit cues within a shared visual context for resolving deictic referencing and precise intent definition.

\subsection{Ethical and Privacy Considerations for Always-On Wearable AI}

User Privacy. Continuous first-person recording captures visual field, sensitive intent and behavioral data. This data poses a risk of being misused to predict user behaviors and preferences. Our system provides clear status visibility of recording indicators and enables users to interrupt the process at any time via physical switches or interface controls.  We have implemented strict storage and permission protocols: the current system stores only inference JSON text fields and image frames related to interaction objects, retaining no user identity information and strictly prohibiting unauthorized access. Future implementations could prioritize on-device processing to reduce data leakage risks.

Bystander Privacy. Wearable cameras risk recording bystanders without consent, inadvertently capturing faces or private moments. Our participants emphasized that their concerns are highly context-dependent.
They were comfortable sharing their view during instrumental tasks with the coffee machine, but expressed strong resistance during intimate family activities, such as playing with family members or building blocks with children.
Technical solutions must ensure bystander awareness, such as prominent LEDs or audio feedback to signal AI activity~\cite{bhardwaj2024focus}.
Future systems should implement automatic filtering to disable video streams when third-party faces are detected.

\subsection{Limitations and Future Work}


About system design, given the computational costs of VLM continuous parsing of video frames, our prototype system employs static triggering rules using a YOLO model to balance responsiveness. However, constrained by the inherent processing latency of VLM inference, the system retains a delay of 4-5 seconds. 
While participants found the latency acceptable for the evaluated tasks, they noted that such delays could become problematic in time-sensitive scenarios, such as safety warnings, cooking, or fluid social interactions.
Furthermore, this rule-based approach lacks flexibility.
For future work, we plan to collect a large-scale first-person dataset to train specialized lightweight vision models that can detect user intent with low latency at the edge. In addition, some participants suggested presenting brief 'thinking indicators' during inference delays to signal system activity and alleviate anxiety while waiting.

About study limitations, first our participants~($N=60$), recruited via social media, consisted primarily of younger individuals with high AI acceptance. And our user study was conducted in a controlled laboratory setting.
Future work should involve in-the-wild studies to evaluate its applicability across broader populations.
Second, while our study offers a preliminary exploration regarding the system $\times$ task interaction effects of cognitive alignment in the first-person perspective, we acknowledge that a larger sample size is required to ensure the robustness of these findings. Future work should conduct large-scale empirical investigations to establish generalized implications.
Finally, the ablation study employed an offline simulation to isolate the impact of user interaction. This design decouples the system from users’ real-time reactions. Also, the memory update~(component II) and reflection~(component III) modules are functionally interdependent, we evaluated them jointly. 
Future work should conduct online ablation studies to capture dynamic behavioral adaptation, and explore ways to disentangle these components without introducing additional confounds.

Beyond specific tasks, \textit{Eye2Eye} demonstrates significant potential for domain transfer.
Since \textit{Eye2Eye} builds common ground purely through real-time user interaction without pre-programmed knowledge, it can adapt to diverse fields, such as industrial assembly~(for error detection) or assistive technology for populations with cognitive challenges~(e.g., children, older adults).
Many participants~(P5, P8, P16) noted that for users who have limited prior knowledge or process information more slowly, a system that can patiently observe, learn from scratch, and provide personalized visual assistance could be truly empowering, helping them better understand and interact with the world around them.
Technically, the \textit{Eye2Eye} framework supports extension to non-AR or lower-fidelity wearable contexts by adapting specific input/output components without restructuring the pipeline.
Future iterations could adapt the joint attention component to lightweight smart glasses by leveraging head-pose estimation or explicit pointing cues captured by the egocentric camera.
Our current prototype employs the Apple Vision Pro to obtain high-fidelity visual copresence. However, the situated feedback component can extend to non-AR hardware such as earbuds and smart watches by utilizing spatial audio or haptics to maintain the shared context.



\section{Conclusion}

This study explores a method to bridge the communication and understanding gulfs that hinder fluent human-AI collaboration. We presented the \textit{Eye2Eye} framework, which transforms the first-person perspective into a shared perception channel. This framework, comprised joint attention, accumulated common ground, and reflective situated feedback, enables a bidirectional loop for cognitive alignment between humans and AI. Through a prototype implementation and a controlled user study, we demonstrated that this approach significantly reduces interaction friction, improves task performance, and enhances copresence and trust. Findings from post-hoc pipeline evaluation confirmed the strong synergy of the \textit{Eye2Eye} pipeline. 
Ultimately, we further reflect on the power of the shared first-person channel and the design implications brought about by several unexpected tensions, thereby redefining human-AI partnerships as a unified joint cognitive system that shares perception and understanding to achieve common goals.




\begin{acks}
This research was supported by the National Natural Science Foundation of China under Grant No. 62502436, and by the Zhejiang Provincial Natural Science Foundation of China under Grant No. LMS26F020004.
\end{acks}

\bibliographystyle{ACM-Reference-Format}
\bibliography{sample-base}

\appendix

\clearpage
\section{Appendix: Configurations of Different Pipelines}

\begin{table*}[b]
  \caption{Configurations of full \textit{Eye2Eye} pipeline~(ours) and the ablation variants.}
  \label{tab:ablation-config}
  \centering
  \footnotesize
  \renewcommand{\arraystretch}{1.3} 
  
  \begin{tabularx}{\textwidth}{l X X X X}
    \toprule
    \textbf{Method} & \textbf{Ours} & \textbf{w/o-JA} & \textbf{w/o-CG-SF} & \textbf{w/o-JA-CG-SF~(baseline)} \\
    \midrule
    
    \textbf{Memory} &
    Can access and update object cards &
    Can access and update object cards &
    Cannot access or update object cards &
    Cannot access or update object cards \\
    \midrule
    
    \textbf{Attention} &
    Uses joint attention alignment &
    Removes joint attention alignment, but still uses target bounding boxes and object nodes to maintain object-level context &
    Uses joint attention alignment &
    Removes joint attention alignment \\
    \midrule
    
    \textbf{Input Stream} &
    Context, image frames, bounding boxes, content of object nodes, memory updates &
    Context, image frames, bounding boxes, content of object nodes, memory updates &
    Context, image frames, bounding boxes &
    Context, image frames, bounding boxes \\
    \midrule
    
    \textbf{Trigger Threshold} &
    Fixates on the same object for $\ge $ 6 seconds,, or triggers when overlap threshold 0.85 is reached &
    Fixates on the same object for $\ge $ 6 seconds, or triggers when overlap threshold 0.85 is reached &
    None &
    None \\
    \midrule
    
    \textbf{Feedback Strategy} &
    Produces situational intention predictions, multimodal suggestion prompts, and memory-update feedback &
    Produces situational intention predictions, multimodal suggestion prompts, and memory-update feedback &
    Produces only situational intention predictions and multimodal suggestion prompts &
    Produces only situational intention predictions and multimodal suggestion prompts \\
    \midrule
    
    \textbf{Output} &
    Updates object cards according to memory-update feedback; logs process data; sends multimodal suggestions to the frontend &
    Updates object cards according to memory-update feedback; logs process data; sends multimodal suggestions to the frontend &
    Logs process data; sends multimodal suggestions to the frontend &
    Logs process data; sends multimodal suggestions to the frontend \\
    \midrule
    
    \textbf{AI Model} &
    Video understanding: \textit{gemini-2.5-flash-preview-native-audio-dialog}; VLM: \textit{gpt-4o} &
    Video understanding: \textit{gemini-2.5-flash-preview-native-audio-dialog}; VLM: \textit{gpt-4o} &
    Video understanding: \textit{gemini-2.5-flash-preview-native-audio-dialog}; VLM: \textit{gpt-4o} &
    Video understanding: \textit{gemini-2.5-flash-preview-native-audio-dialog}; VLM: \textit{gpt-4o} \\
    
    \bottomrule
  \end{tabularx}
\end{table*}

To ensure fair comparison and clarify the configuration of each ablation and baseline condition, we summarize the input streams, model versions, trigger strategies, feedback strategies and output for all methods in Table~\ref{tab:ablation-config}.
In the user study, the baseline configuration is identical to the w/o-JA-CG-SF.
Both remove all first-person cognitive alignment components, and only retain the foundational models. 

For the ablation study, all conditions operate on the same first-person video stream and audio input captured from the AR headset. We also use the same video-understanding model and visual language model, and run on the same server and hardware configuration.




\end{document}